\documentclass[fleqn,usenatbib]{mnras}

\usepackage{newtxtext,newtxmath}
\usepackage[T1]{fontenc}

\DeclareRobustCommand{\VAN}[3]{#2}
\let\VANthebibliography\thebibliography
\def\thebibliography{\DeclareRobustCommand{\VAN}[3]{##3}\VANthebibliography}

\usepackage{graphicx}	% Including figure files
\usepackage{amsmath}	% Advanced maths commands
\usepackage{xcolor}
\usepackage{physics}
\usepackage{subcaption}
\usepackage{soul}
\title[Quality over Quantity]{Quality over Quantity: Optimizing pulsar timing array analysis for stochastic and continuous gravitational wave signals}

\author[]{
Lorenzo Speri$^{1}$, \thanks{lorenzo.speri@aei.mpg.de} 
Nataliya~K.~Porayko$^{2}$, 
Mikel Falxa$^{3}$, 
Siyuan~Chen$^{4}$, 
Jonathan~R.~Gair$^{1}$, 
Alberto~Sesana$^{5,6}$,
\newauthor
Stephen~R.~Taylor$^{7}$
\\
$^{1}$Max Planck Institute for Gravitational Physics (Albert Einstein Institute), Am M\"{u}hlenberg 1, 14476 Potsdam, Germany\\
$^{2}$Max-Planck-Institut f\"{u}r Radioastronomie, Auf dem H\"{u}gel 69, 53121 Bonn, Germany\\
$^{3}$Université de Paris, CNRS, Astroparticule et Cosmologie, 75013 Paris, France\\
$^{4}$Kavli Institute for Astronomy and Astrophysics, Peking University, Beijing 100871, P. R. China\\
$^{5}$ Dipartimento di Fisica ``G. Occhialini'', Universit\`{a} degli Studi di Milano-Bicocca, Piazza della Scienza 3, I-20126 Milano, Italy\\
$^{6}$ INFN, Sezione di Milano-Bicocca, Piazza della Scienza 3, 20126 Milano, Italy\\
$^{7}$Department of Physics \& Astronomy, Vanderbilt University, 2301 Vanderbilt Place, Nashville, TN 37235, USA\\
}

\date{Accepted XXX. Received YYY; in original form ZZZ}

\pubyear{2022}

\begin{document}
\label{firstpage}
\pagerange{\pageref{firstpage}--\pageref{lastpage}}
\maketitle

\begin{abstract}
The search for gravitational waves using Pulsar Timing Arrays (PTAs) is a computationally expensive complex analysis that involves source-specific noise studies. As more pulsars are added to the arrays, this stage of PTA analysis will become increasingly challenging. Therefore, optimizing the number of included pulsars is crucial to reduce the computational burden of data analysis. Here, we present a suite of methods to rank pulsars for use within the scope of PTA analysis. First, we use the maximization of the signal-to-noise ratio as a proxy to select pulsars. With this method, we target the detection of stochastic and continuous gravitational wave signals. Next, we present a ranking that minimizes the coupling between spatial correlation signatures, namely monopolar, dipolar, and Hellings \& Downs correlations. Finally, we also explore how to combine these two methods. We test these approaches against mock data using frequentist and Bayesian hypothesis testing. For equal-noise pulsars, we find that an optimal selection leads to an increase in the log-Bayes factor two times steeper than a random selection for the hypothesis test of a gravitational wave background versus a common uncorrelated red noise process. For the same test but for a realistic EPTA dataset, a subset of 25 pulsars selected out of 40 can provide a log-likelihood ratio that is 89\% of the total, implying that an optimally selected subset of pulsars can yield results comparable to those obtained from the whole array. We expect these selection methods to play a crucial role in future PTA data combinations.

\end{abstract}

\begin{keywords}
Gravitational Waves -- Data Analysis -- Pulsars
\end{keywords}

%%%%%%%%%%%%%%%%%%%%%%%%%%%%%%%%%%%%%%%%%%%%%%%%%%

%%%%%%%%%%%%%%%%% BODY OF PAPER %%%%%%%%%%%%%%%%%%
%----------------------------------------------------------
\section{Introduction}
%----------------------------------------------------------

Pulsar Timing Array (PTA) experiments search for nanohertz-frequency gravitational waves (GWs) through induced shifts in radio-pulse arrival times from Galactic millisecond pulsars \citep{sazhin-1978,detweiler-1979}. The timing precision and regularity of the pulse times of arrival (TOAs) from these pulsars make them exquisite laboratories for studying a variety of astrophysical and fundamental physics phenomena \citep[e.g.,][]{2009MNRAS.400..951V}. This includes GWs, which impart changes to the proper separation of Earth and the pulsar, causing pulses to arrive earlier or later than expected. These timing deviations are a function of the GW source characteristics, as well as the geometry of the GW source relative to the Earth-pulsar line-of-sight. Upon fitting a deterministic timing ephemeris (describing leading order behavior such as the rotational period, spindown rate, etc.) to a pulsar's TOAs, the remaining timing residuals can be analysed to search for the presence of GW signals amidst noise contributions. In a single pulsar's timing residuals, GW signals can easily be conflated with intrinsic pulsar noise effects \citep[e.g.,][and references therein]{2010ApJ...725.1607S} or even poorly understood artifacts of the ionized interstellar medium that radio pulses must traverse \citep[e.g.,][and references therein]{2010arXiv1010.3785C}. But by constructing an array of pulsars, the fact that the GW-induced timing deviations are correlated between pulsars can be leveraged to distinguish it from uncorrelated astrophysical and instrumental noise processes \citep{1990ApJ...361..300F}.

Several large collaborations have been monitoring ensembles of millisecond pulsars over long timing baselines in a bid to detect both a stochastic GW background (GWB) and individually-resolvable GW sources. These include the European Pulsar Timing Array \citep[EPTA,][]{2013CQGra..30v4009K}, the North American Nanohertz Observatory for Gravitational waves \citep[NANOGrav,][]{2013CQGra..30v4008M}, and the Parkes Pulsar Timing Array \citep[PPTA,][] {2013PASA...30...17M}. Together with the more recently established Indian PTA \citep[InPTA,][]{inpta}, these collaborations constitute the International Pulsar Timing Array \citep[IPTA,][]{2016MNRAS.458.1267V,Perera:2019sca}, which aims to synthesize the aforementioned regional efforts to achieve more significant and rapid discoveries. Other recent timing efforts include the Chinese PTA \citep[CPTA,][]{cpta}, the MeerTIME program \citep{meertime} conducted at the MeerKAT telescope \citep{Camilo:2018hsu}, CHIME/Pulsar \citep{chime}, GMRT \citep{1990IJRSP..19..493S} and FAST \citep{FAST}. Recent results from NANOGrav \citep{2020ApJ...905L..34A}, the PPTA \citep{2021ApJ...917L..19G}, the EPTA \citep{chen2021}, and the IPTA \citep{IPTA2022} all show strong evidence in favor of a common-spectrum process versus independent red-noise processes with Bayes factors of order $\sim 10^3-10^4$.
These stochastic processes have similar spectral characteristics with estimated amplitudes around $A \sim2-3\times 10^{-15}$, and are all in broad agreement with expectations for a GWB generated by an astrophysical population of supermassive black-hole binaries \citep[SMBHBs, e.g.,][]{2021MNRAS.502L..99M}. However, there is not yet significant evidence for the distinctive pattern of inter-pulsar correlations, known as the Hellings \& Downs (HD) curve. {In fact such evidence needs more time to emerge than the presence of a common process \citep{NANOGrav:2020spf,Romano:2020sxq}.}

Building evidence for GW-induced inter-pulsar correlations requires many well-timed pulsars in order to forge effective pairings across different angular separations in order to trace out the HD pattern \citep{hd_paper}. This pattern is mostly quadrupolar in angular separation, with two zero crossings between $0^\circ$ and $180^\circ$.  Yet there are several issues associated with building an effective pulsar array for GW detection. $(i)$ First, we are constrained by the Galactic distribution of millisecond pulsars, so there is little reason to consider array geometries that contradict this. $(ii)$ Furthermore, if one were to only try to discover new pulsars that would maximize the significance of HD correlations, then the best strategy would be to survey close to the most sensitive pulsars. However, this would not trace the full pattern of this correlation curve, thereby severely inhibiting our ability to discriminate it from systematic noise processes that can also induce inter-pulsar correlations \citep{2016MNRAS.455.4339T}. The latter include solar-system ephemeris errors that create dipolar correlations \citep{2020ApJ...893..112V,2010ApJ...720L.201C,2019MNRAS.489.5573G,2018MNRAS.481.5501C,Roebber_2019}, and long-timescale systematics in time standards that create monopolar correlations \citep{2012MNRAS.427.2780H,2020MNRAS.491.5951H}. $(iii)$ Finally, the next generation of radio facilities such as DSA-2000 \citep{Hallinan2021DSA}, the Square Kilometre Array \citep[SKA,][]{dewdney2009square,2015aska.confE..37J}, and the next-generation Very Large Array \citep[ngVLA,][]{murphy2018science} will lead to a torrent of new pulsars and observations. Future PTA data analysts will need metrics to judge which pulsars will most effectively characterize the GWB and resolve multiple individual SMBHBs out of this confusion background.

Therefore, exploring how to optimize the observing and analysis strategies of PTA experiments is crucial.
In previous works, computational techniques to optimize the observational schedule \citep{2012MNRAS.423.2642L,2018ApJ...868...33L}, and arrival-time precision as a function of radio frequency and bandwidth \citep{Lam_2018} have been investigated. 
In \cite{Roebber_2019}, the author proposed a technique to optimize the disentangling between different spatial correlations and, therefore, to separate the signal due to GWs from that produced by clock or ephemeris errors. This paper also argued that such a method could be used to decide which pulsars should be included in PTAs. Beyond standard quality checks related to a pulsar's long-term timing stability, PTA searches aim to include as many pulsars as possible. However, a standard timing baseline cut of $\sim 3$ years is usually made in order to reduce the data volume while at the same time ensuring that all pulsars inform GW frequencies $\lesssim10$~nHz where a GW background signal should be strongest.

In this work, we introduce for the first time a robust methodology for pulsar selection optimization in order to detect and characterize both the stochastic background and single continuous gravitational wave (CGW) sources.
We develop ranking (or selection) methods to understand which pulsars contribute most to GW searches, where we target three key analyses: $(i)$ detection of a GWB versus a Common Uncorrelated Red Noise (CURN) process, $(ii)$ detection of a GWB versus Monopolar and Dipolar correlated signals, $(iii)$ detection of CGW sources.
These methods use statistical tools introduced in previous studies, making our methods easily implemented within established pipelines. Each method takes as input the intrinsic timing and noise properties of the whole pulsar array -- which could be potentially provided by previous data releases-- and outputs a ranked list of pulsars for a specified GW search.

This paper is organised as follows. We review the standard PTA statistical tools such as likelihood and frequentist and Bayesian hypothesis testing in Section~\ref{subsec:pta_likelihood} and \ref{subsec:hyp_testing}. These tools are used to test the performance of the ranking methods introduced in Section~\ref{subsec:ranking_stochastic} and \ref{subsec:continuous_snr}.
In particular, the ranking method based on signal-to-noise ratio (SNR) maximization is presented in \ref{subsubsec:snr_max}, and the one aimed at disentangling different spatial correlations in \ref{subsubsec:decoupling_mat}. 
In Section~\ref{subsec:continuous_snr} we develop a selection method that targets the search for continuous gravitational wave signals.
The results are presented in Section~\ref{sec:results} where the selection methods are tested using simulated datasets with increasing level of noise complexity.
We conclude with our expectations for future investigations in Section \ref{sec:conclusion}.

%----------------------------------------------------------
\section{Methods}
\label{sec:methods}
%----------------------------------------------------------

%----------------------------------------------------------
\subsection{Pulsar Timing Array Likelihood}
\label{subsec:pta_likelihood}
%----------------------------------------------------------
In this section we introduce the marginalized PTA likelihood which is ultimately the fundamental tool for the statistical analysis of PTA data \citep{2009MNRAS.395.1005V}. We predominantly follow the ``Gaussian process'' treatment described in details in \citet{2014PhRvD..90j4012V, 2016ApJ...821...13A}.
The TOAs for each pulsar can be represented by a vector $\vec{t}$ of length $N_{\text{TOA}}$. $\vec{t}$ can be written as a sum of a deterministic and a stochastic component: $\vec{t} = \vec{t}_{\text{det}} + \vec{t}_{\text{sto}}$. 

The deterministic part comprises the so-called timing model which depends on a set of timing parameters $\vec{\beta}$. The timing model describes the intrinsic spin evolution of a source, propagation effects as well as time delays associated with the relative motion of a source and the Earth and kinematic and light propagation effects in the binary system \citep[see e.g.][]{2012hpa..book.....L}. The initial estimate of the $m$ timing model parameters $\vec{\beta}_0$ is obtained using the minimization of the sum of the squares of the residuals $\delta \vec{t} = \vec{t} - \vec{t}_{\text{det}} (\vec{\beta})$. This least-square linear fit to the timing model, which is performed using the \texttt{TEMPO2} software \citep{tempo2_2, tempo2_3}, is equivalent to likelihood maximization when assuming Gaussian white noise errors. In reality the stochastic noise component is dominated by colored noises. Assuming that the initial estimate of the timing parameters $\vec{\beta}_0$ obtained from \texttt{TEMPO2} does not differ significantly from the final estimate $\vec{\beta}_f$ obtained from a full analysis that includes more sophisticated stochastic noise modelling, the timing model can be approximated to impact the timing residuals linearly via the term $\vb{M}\vec{\epsilon}$, where $\vec{\epsilon} = \vec{\beta}_f-\vec{\beta}_0$ and $\vb{M}$ is an $N_{\text{TOA}} \times m$ \textit{design matrix} \citep{2009MNRAS.395.1005V}.

The correlated components of the stochastic piece $\vec{t}_{\text{sto}}$ are modelled in terms of a Fourier decomposition \citep{2013PhRvD..87j4021L}. In practice, the analysis focuses on the noise with dominant power at lower frequencies, so that only a finite number of Fourier components $N_f$ are used. In this case the signal can be written in a matrix form of the type $\vb{F}\vec{a}$, where the vector $\vec{a}$ of length $2 N_{\text{freqs}}$ contains the Fourier coefficients, whereas the $N_{\text{TOA}} \times 2N_{\text{freqs}}$ matrix $\vb{F}$ is constructed with alternating columns of sines and cosines evaluated at the TOAs of each pulsar. The base sampling frequency is given by the inverse of the observation timespan of the entire pulsar timing array, $1/T$. 

The influence of white-noise on the timing residuals is described by the $N_{\text{TOA}} \times N_{\text{TOA}}$ white noise covariance matrix $\vb{N}$. Finally, the noise-mitigated timing residuals $\vec{r}$, which is our best approximation to the white noise $\vec{n}$ for each pulsar can be written in a compact form as a function of the input residuals $\delta \vec{t}$:
\begin{equation}
    \vec{r} = \delta \vec{t} - \vb{T} \vec{b} \qquad \vb{T} = \mqty[ \vb{M} & \vb{F}] \qquad \vec{b} = \mqty[ \vec{\epsilon} \\ \vec{a}] \,,
\end{equation}
and the likelihood is given by:
\begin{equation}
    p({ \vec{\delta t}} | \vec{b}) =
    \frac{
        \exp{-\frac{1}{2}  \vec{r}^\text{T} \, \vb{N}^{-1} \, \vec{r}}
        }{
        \sqrt{2 \pi \det{\vb{N}}} 
        }
        \,.
\end{equation}
The prior covariance and corresponding Gaussian prior on the coefficients $\vec{b}$ are written as:
\begin{equation}
    \vb{B} = \mqty[\boldsymbol{\infty} & \vb{0} \\ \vb{0} & \boldsymbol{\phi} ] 
    \qquad
    p({ \vec{b}} | \vec{\phi}) =
    \frac{
        \exp{-\frac{1}{2}  \vec{b}^\text{T} \, \vb{B}^{-1} \, \vec{b}}
        }{
        \sqrt{2 \pi \det{\vb{B}}}
        }
         \,,
\end{equation}
so that the timing model piece of $\vec{b}$ is a uniform unconstrained prior on the timing model parameters $\vec{\epsilon}$, and the spectrum of all low-frequency processes enters in the variance $\boldsymbol{\phi}$ as:
\begin{equation}
    \phi_{(ai),(bj)} = \Gamma_{ab} S_i \delta_{ij} + P_{a i} \delta_{ab} \delta_{ij}  \,.
\end{equation}
where the intrinsic low-frequency (``spin-noise") spectrum of pulsar $a$ at the $i^{\text{th}}$ sampling frequency is represented by $P_{a i}$, and the GWB spectrum, which is common to all pulsars, is given by $S_i$. 
Both of these processes can be modelled with a power-law functional form:
\begin{equation}
    P_{ai} = \frac{A_{a} ^2 }{12 \, \pi^2 T} \, \qty(\frac{f_i}{\text{yr}^{-1}})^{-\gamma_{a}} \text{yr}^2 \,.
\end{equation}
The reduction in correlated power due to the spatial separation of the pulsars is described by the overlap reduction function (ORF) $\Gamma_{ab}$ between pulsars $a$ and $b$. For an isotropic and stochastic GWB, the ORF is described by the HD curve \citep{hd_paper}, which depends only on the angular pulsar separation.
If we group all the red noise and GWB spectral hyper-parameters into the vector $\vec{\eta}$ we can obtain the  likelihood of the full PTA array \citep{2014PhRvD..90j4012V}, marginalized over $\vec{b}$:
\begin{equation}
\begin{split}
    \mathcal{L}(\vec{\eta}) &= p(\{\delta\vec{t} \}|\vec{\eta}) = \int \prod_{a=1} ^{N} p(\delta\vec{t}_a|\vec{b}_a) \times p(\{\vec{b}\}|\vec{\eta}) d^{N} \vec{b} \, ,\\
    \ln \mathcal{L}&=
        - \frac{1}{2} \qty[\delta\vec{t}^\text{T} \, \vb{C}^{-1} \, \delta\vec{ t} +  \Tr \ln{2 \pi \vb{C}} ] \,,
\end{split}
\label{eq:likelihood}
\end{equation}
where $\vb{C} = \vb{N}+\vb{T} \vb{B} \vb{T}^{\text{T}}$, and $N$ is the total number of pulsars.
A deterministic signal $\vec{s}(\vec{\theta})$ can be incorporated in the modelling by performing the following replacement $\delta \vec{t} \rightarrow \delta \vec{t} - \vec{s}(\vec{\theta})$. More details on likelihood construction and handling correlated noise processes in pulsar timing analysis can be found in e.g., \citet{2013MNRAS.428.1147V, 2015ApJ...810..150A, 2016ApJ...821...13A, 2021arXiv210513270T}.

Having constructed the PTA marginalized likelihood, we can estimate the parameters $ \vec{\eta}$. In frequentist inference, the true model parameters are considered to be fixed $\eta _{ \textrm{\tiny True}}$, and are estimated by maximizing the likelihood to obtain the maximum likelihood estimator (MLE), $\vec{\eta}_{\text {\tiny MLE} }$. In Bayesian inference, model parameters are no longer regarded as fixed, but are themselves random variables. The probability distribution of the parameter values before the data acquisition (the prior distribution $p(\vec{\eta})$) is updated to a probability distribution after the data incorporation (the posterior distribution $p(\vec{\eta}|\delta\vec{t})$) through the likelihood of the observed data $\mathcal{L}(\delta\vec{t} | \vec{\eta})$. With several intrinsic noise parameters per pulsar, in addition to several global parameters describing the GW signal, the posterior distribution can be as high as $\mathcal{O}(100)$-dimensional. Thus, it is typically explored and sampled numerically using Markov chain Monte Carlo (MCMC) techniques.
%----------------------------------------------------------
\subsection{Hypothesis testing}
\label{subsec:hyp_testing}
%----------------------------------------------------------
The essential step of the PTA analysis is testing whether the observed data are consistent with our expectations, e.g. the presence of a GW signal or its absence. Therefore, we use hypothesis testing to investigate if the data provides sufficient evidence for one hypothesis $\mathcal{H}_1$ with respect to another one $\mathcal{H}_2$. The tools developed in this section will be used in Sec.~\ref{sec:results} as a proxy to test our selection methods.

If we adopt a frequentist approach, we can maximize the likelihood under each hypothesis to find the MLE for the parameters, i.e., $\vec{\eta}_{\text {\tiny MLE 1} } = \text{max}_{\vec{\eta}} \ln \mathcal{L}(\vec{\eta}| \mathcal{H}_1) $ and analogously for $\mathcal{H}_2$.
Then, the log-likelihood ratio defined as:
\begin{equation}
    \ln{\Lambda} =  \ln \mathcal{L}( \vec{\eta}_{\text {\tiny MLE 1} } | \mathcal{H}_1) - \ln \mathcal{L}( \vec{\eta}_{\text {\tiny MLE 2} } |\mathcal{H}_2)
\end{equation}
can be used to test whether our data supports hypothesis $\mathcal{H}_1$ with respect to $\mathcal{H}_2$. Roughly speaking, a large value of $\ln \Lambda$ indicates a stronger support for $\mathcal{H}_1$ with respect to $\mathcal{H}_2$. Therefore, we can use $\ln \Lambda$ to assess if an optimally selected subset of pulsars supports our expectations as much as the full dataset.

To statistically quantify the significance of a measured log-likelihood value it is necessary to create multiple realizations of the data under the reference hypothesis $\mathcal{H}_2$. For each realization, we must then evaluate the log-likelihood ratio to obtain a distribution of $\ln \Lambda$ under the reference hypothesis. This distribution can be used to calculate the p$-$value of the measured log-likelihood. This approach is only viable if our ranking methods are tested on mock dataset realizations.

In reality, we cannot generate multiple realizations of the data because we do not have access to the true parameters and data generation process. We have access only to the most likely values of such parameters from previous data releases. Therefore, we can use those for the data generation of the reference hypothesis. By evaluating the p-value for the real dataset, we estimate the significance of such an experiment and check the consistency of our assumptions on the data generation process. Similar tests are extensively used in PTA analysis \citep[see sky scrambles, phase shifts, and optimal statistic analysis, e.g.,][]{Chamberlin_2015,2016PhRvD..93j4047C,Taylor:2016gpq}. We evaluate this procedure as a consistency check for hypothesis testing of a realistic PTA analysis in Sec.~\ref{subsec:optimize_epta}.

In Bayesian statistics, the Bayes Factor (BF)
\begin{equation}
    \text{BF} = \frac{
     \int \dd \vec{\eta} \mathcal{L}(\delta\vec{t}\, | \vec{\eta}, \mathcal{H}_1) p(\vec{\eta}, \mathcal{H}_1)
    }{
     \int \dd \vec{\eta} \mathcal{L}(\delta\vec{t}\, | \vec{\eta}, \mathcal{H}_2) p(\vec{\eta}, \mathcal{H}_2)
    }
\end{equation}
is used to assess which model is favored by the observations, assuming that the two models are equally probable a priori. A ``rule of thumb'' for interpreting Bayes’ factors is presented in \cite{kassBayesFactors1995}, where $\text{BF}>20$ is considered strong evidence for $\mathcal{H}_1$.\footnote{Alternatively, the distribution of the Bayes factor can be computed under the null hypothesis and used, in a frequentist way, to produce a mapping between p-values and Bayes factors. However, this approach is computationally expensive.}

{If the posterior volumes of the two hypotheses are approximately the same, then the log-likelihood ratio at the MLE is approximately equal to the log-Bayes factor, i.e. $\ln \text{BF} \approx \ln \Lambda$ \citep{2017LRR....20....2R,NANOGrav:2020spf}.}

In practice, BFs are widely used to perform robust statistical analysis, including hypothesis testing, when processing real PTA datasets. In this work, full Bayesian inference is only used for computationally feasible analysis of simplified datasets. For the realistic mock datasets which require more sophisticated noise modelling, we utilise the log-likelihood ratio test as it requires fewer computational resources.

%----------------------------------------------------------
\subsection{Ranking pulsars for stochastic signal searches}
\label{subsec:ranking_stochastic}
%----------------------------------------------------------
One of the primary goals of the current PTA experiments is to detect the stochastic GWB from a population of SMBHBs. 
An isotropic GWB manifests itself as a long timescale, low-frequency (or red) common signal across the pulsars in a PTA. This common signal is characterized by the common spectrum and the inter-pulsar spatial correlations.
The distinctive signature of the gravitational nature lies in this correlation which depends only on the pulsar's angular separation and has an expectation value given by the HD curve \citep{hd_paper}.
Current experiments found strong evidence for the presence of a common red noise signal. While such a signal could potentially represent the expected GWB from SMBHBs, there is not yet strong evidence for either HD or other alternative angular correlations.

Motivated by these latest results, in Sec. \ref{subsubsec:snr_max} we design a method to identify the optimal subset of pulsars for increasing the confidence in the detection of an HD correlation, whereas in Sec. \ref{subsubsec:decoupling_mat} we use the decoupling formalism to find the best subset of pulsars for distinguishing this correlation from alternative hypotheses. {Recent work has cautioned that GWB upper limits can be biased and even lie below the true value when small ($\lesssim 20$) combinations of pulsars are analyzed \citep{2022ApJ...932..105J}. Our work here is likely immune from such unwanted effects for several reasons: $(i)$ the field of PTAs has moved beyond the regime of setting upper limits, to now estimating the statistical parameters of a common process and performing model selection on spatial correlations; and $(ii)$ our metrics here are based on the detectability and discrimination of stochastic processes, rather than upper limits.}   
%----------------------------------------------------------
\subsubsection{Spatially correlated signal-to-noise ratio maximization}
\label{subsubsec:snr_max}
%----------------------------------------------------------
As previously mentioned, the target signal is described by a correlated red noise process $S(f)$ with spatial correlations $\Gamma _{ab}$. An optimal subset of pulsars can be constructed based on an optimal statistic that maximizes the detection probability at a fixed false alarm probability for this specific case. As a proxy for this, it is convenient to consider statistics that maximize the signal-to-noise ratio (SNR), which is the ratio of the expected value of a statistic in the presence of a signal, $\mu_1$, to its standard deviation. The standard deviation can either be computed in the absence of a signal, $\sigma_0$, or in the presence of a signal, $\sigma_1$. In \cite{Rosado_2015}, the authors introduce two statistics: the A-statistic constructed by maximizing $\mu_1/\sigma_0$ and the B-statistic constructed by maximizing $\mu_1/\sigma_1$. This procedure leads to the respective SNR definitions:
\begin{align}\label{eq:signal-to-noise-A}
    \text{SNR}_\text{A}^2  &= 2\sum_{a>b} \int \frac{
    \Gamma^2 _{ab} \, S^2 (f) \, T_{ab}
    }{
    P_a (f) P_b (f)
    } \dd f\, , 
    \\
    \text{SNR}_\text{B}^2  &= 2\sum_{a>b} \int \frac{
    \Gamma^2 _{ab} \, S^2 (f) \, T_{ab}
    }{
    [P_a (f) + S (f)] [P_b (f) + S (f)] + S^2(f)  \Gamma^2 _{ab}
    } \dd f\, . \label{eq:signal-to-noise-B}
\end{align}
We use these quantities as a proxy to identify the best subset of pulsars from the full array.
SNR$_\text{A}$ and SNR$_\text{B}$ are obtained under the expectation value of the true hypothesis and do not depend on the timing residuals but only on the general properties of the pulsars' red and white noises. In Eq.~(\ref{eq:signal-to-noise-A}-\ref{eq:signal-to-noise-B}), the sum is over the pulsar pair $a,b$, with $a > b$ and $T_{ab}$ is the overlapping time of observation of the $a,b$ arrays. The term $P_a (f)$ represents the sum of the intrinsic noise processes of pulsar $a$ such as red noise, white noise, etc. :
\begin{equation}
\begin{split}
    P_a(f) &= P_{\text{rn}} + P_{\text{wn}} + ... \\
    &= \frac{A_{a} ^2}{12 \pi^2} \qty(\frac{ f}{ \text{yr}^{-1} })^{-\gamma_a} \, \text{yr}^3 + 2 \sigma ^2 \Delta t + ...
\end{split}
\end{equation}
where $\sigma$ is the root-mean-square (RMS) error and $\Delta t$ is the cadence of the TOAs. We also assume that the correlated noise process $S(f)$ can be described by a power-law functional form.

As pointed out in \cite{Rosado_2015}, the SNR$_\text{B}$ is more robust in the strong-signal regime. In fact, as we can see from Eq.~(\ref{eq:signal-to-noise-A}-\ref{eq:signal-to-noise-B}), one of the useful differences with respect to the other statistic is that SNR$_\text{B}$ does not diverge for $S\gg P_a$. The SNR$_\text{B}$ is very similar to the so-called optimal statistic SNR presented in \citep{Siemens_2013,Chamberlin_2015}, however the last term in the denominator of SNR$_\text{B}$ is missing in those studies.

One downside of using the SNR$_\text{B}$ of Eq.~(\ref{eq:signal-to-noise-B}) is that it assumes the amplitude and slope of $S(f)$ to be known. Since we have constraints on such parameters from the current PTA experiments, we can assume these to be known and use them to calculate the SNR. We will later show that the selection procedure using this SNR is not strongly affected by the variations of these quantities when estimated over noise realizations. The SNR$_\text{A}$ definition has the advantage that the amplitude factors out and therefore its maximization is not affected by the choice of $A_\text{GWB}$.

In theory, we would need to compare the SNRs with all possible combinations of subsets of pulsars from the whole array. Since this is computationally intractable in practice, we start from a few fiducial pulsars and add pulsars one by one until we reach the desired level of SNR. {We will see in Sec.~\ref{subsec:optimize_epta} that this ``one-by-one'' implementation of SNR-maximization performs very well, reaching a high proportion of the full data set BF with only a small selection of pulsars. The small improvement that might be achieved from an exhaustive search of all possible pulsar subsets is unlikely to be worth the considerable increase in computational cost.}

If we set the spatial correlation $\Gamma_{ab}$ to be the HD correlation, we can use these SNRs to rank pulsars and increase the detection probability of a GWB. Therefore, the SNR-maximization selection method introduced here aims at providing the best pulsars for the hypothesis test of an HD correlation (hypothesis $\mathcal{H}_1$) versus a CURN (hypothesis $\mathcal{H}_2$).

%----------------------------------------------------------
\subsubsection{Maximization of the decoupling between spatial correlations}
\label{subsubsec:decoupling_mat}
%----------------------------------------------------------
An unambiguous detection of a GWB relies on the characterization of the angular correlation between pulsars. In order to claim a detection, PTA experiments must provide strong evidence that an HD correlation is clearly identified in the data. However, the detection of a GWB is complicated by the presence of other types of correlated signals. Specifically, errors in clocks used to calibrate timing residuals, and poorly determined solar system ephemeris induce large-scale correlations between pulsars and can mimic the effects of a GWB. The irregularities in terrestrial time standards produce signals with monopolar spatial correlation \citep{2012MNRAS.427.2780H, 2020MNRAS.491.5951H}, while emphemeris errors can result in dipolar signals \citep{2010ApJ...720L.201C, 2016MNRAS.455.4339T}. In order to provide an optimal separation of the quadrupole GWB signal from those produced by clock or ephemeris errors, \cite{Roebber_2019} proposed a method to minimize the leakage between spatially correlated noises. We briefly review this formalism here.

The degree to which power from one spatial harmonic can leak into another one can be quantified by the coupling matrix \citep{1973ApJ...185..413P,1994ApJ...430L..89G,2001PhRvD..64h3003W,2002ApJ...567....2H,2002MNRAS.330..405M,2004MNRAS.349..603E}:
\begin{equation}
    K_{(lm),(lm)'} = 
    \int Y_{lm} (\Omega) W(\Omega) Y_{(lm)'} (\Omega) \dd \Omega \, ,
\label{eq:klmlm}
\end{equation}
where $Y_{lm}$ is the spherical harmonic of degree $l$ and order $m$, $W(\Omega)$ is the window function, and the integral is performed over all sky directions, $\Omega$. The Coupling Matrix formalism can be directly applied to the pulsar selection problem. Within the PTA framework, a GWB has maximum power at $l=2$, while clock noise and ephemeris noise appear at $l=0$ and $l=1$, respectively. Therefore, the coupling matrix elements with $l$ from 0 to 2 are of interest for the problem of mode disentangling. While forming an orthonormal basis in the case of continuous coverage ($W(\Omega)=1$ everywhere on the sky), the coupling matrix loses its orthogonality when the sampling of the sky becomes discrete, resulting in non-zero off-diagonal elements in $K_{(l m),(lm)'}$. 

In the context of PTA analysis, the window function is given by the Kronecker-delta modulated by the individual weights $w$ of pulsars placed at sky positions $\hat{p}_a$:
\begin{equation}
W(\Omega)=\sum_a w^a \delta(\Omega-\hat{p}_a) \, .
\label{eq:cp_wght}
\end{equation}
In the case of all-equal pulsars, the choice of the weighting function is straightforward: $w^a=1$ for all pulsars. %in the direction of the source and $w^a=0$, otherwise. 
However, the problem becomes less trivial when each pulsar has different properties (in terms of RMS residuals, observation time, intrinsic red noise, etc.). \cite{Roebber_2019} suggests to use the inverse of the RMS of a source, $1/\sigma^2_a$, as weights, to account for the relative sensitivity of different pulsars in an array. {In order to additionally account for the coloured noise in an array, we will use $\textrm{SNR}_\text{A}\sim 1/\sigma^2_a$ as weights in the coupling matrix formula, where $\textrm{SNR}_\text{A}$ is defined using the self-term ($a=b$) of Eq~(\ref{eq:signal-to-noise-A}).}
Although this is a natural choice, it is worth noting that the optimal choice of the weighting function for the coupling matrix construction does not have a unique solution and in some cases requires a heuristic approach \citep{2004MNRAS.349..603E}. As shown in Appendix \ref{subsec:cm_wgt}, for the two realistic mock datasets described in Sec.~\ref{sec:results}, an $\textrm{SNR}^4_\text{A}$-weighting on average performs better than the other types of weighting function considered. However, in order to provide a definitive solution to the problem of weight selection, extensive testing on more diversified samples of mock datasets is required, which we leave for future work.

The level at which one mode leaks to another is estimated via the ratio of minimum and maximum eigenvalues $\lambda_\textrm{min}/\lambda_\textrm{max}$ of $K_{(l m),(lm)'}$, which is 1 when the coupling matrix is diagonal and drops to 0 when the coupling matrix is ill-defined. Since we are mainly interested in decoupling the spherical harmonics with different $l$, we can average Eq.~(\ref{eq:klmlm}) over $m$. Thus, the final expression for the coupling matrix is \cite{2004MNRAS.349..603E}:
\begin{equation}
M_{l,l'} = \frac{1}{(2l+1)(2l'+1)}\sum_{m, m'}K_{(l, m) (l',m')} \, .
\end{equation}
We construct the pulsar ranking list by selecting those that lead to the largest eigenvalue ratio $\delta_{\lambda}=\lambda_\textrm{min}/\lambda_\textrm{max}$ of the $M_{l,l'}$ matrix.
The Coupling Matrix selection method introduced here aims at providing the best pulsars for the hypothesis test of an HD correlation (hypothesis $\mathcal{H}_1$) versus the presence of all three signals in the data, namely common uncorrelated, monopolar and dipolar spatially correlated red noise processes (hypothesis $\mathcal{H}_2$). 
As pointed out in \cite{Roebber_2019}, the minimum number of pulsars required to disentangle up to $l_\text{max}$ is $\sum_{l=0} ^{l_{\text{max}}} (2l+1) = (l_\text{max}+1)^2$, which is 9 for $l=2$. After averaging over $m$, the coupling matrix $M_{l,l'}$ is well-defined when the number of pulsars is $\geq$3, meaning that at least three pulsars are required to resolve the spatial modes up to the quadrupole. Therefore, when the Coupling Matrix formalism is applied to realistic datasets, in order to avoid ambiguity, the first three pulsars in the ranking are fixed to those with the highest self-SNR. 

\subsubsection{Chimera method: combining SNR- and decoupling-maximization algorithms}
The Coupling Matrix selection method is aimed at disentangling different types of correlations, while the total SNR maximization is disregarded. Therefore, the Coupling Matrix can only be used as a complementary scheme for array optimization, especially, for an array of pulsars in mixed SNR regime\footnote{This means that the vast majority of pulsars in an array are in the weak signal regime \citep{Siemens_2013} and only a few sources actually contain the detectable signal. In this case, the latter are expected to contribute a significant fraction of the whole array sensitivity, while the addition of the former sources is largely irrelevant.}. Here we propose a new selection method that combines the merits of both the Coupling Matrix and SNR maximization: hereafter the ``Chimera''\footnote{The name was inspired by the mythological creature composed of different animal parts. Homer describes it as follows in the Iliad: ``she was of divine stock, not of men, in the fore part a lion, in the hinder a serpent, and in the midst a goat, breathing forth in terrible wise the might of blazing fire.'' \cite{homer2005iliad}} method. The basic idea is to add a new pulsar to a subset, so that the HD-SNR is maximized along with the decoupling power. 
One of the possible norms that satisfies the latter requirement is the multiplication of the relevant scores of both methods, i.e. SNR and eigenvalue ratio:
\begin{equation}
\textrm{SC}_{\textrm{Chimera}}= \textrm{SNR}^2_\textrm{B}\delta_{\lambda}.
\label{eq:chimera_score}
\end{equation}
Note that the ranking of pulsars within the Chimera approach is purely heuristic and the score that we offer in Eq.~(\ref{eq:chimera_score}) is one of many possible choices. As in the case of the Coupling Matrix, the first three pulsars are selected according to the highest self-SNR, while the following ones are picked so that the score in Eq.~(\ref{eq:chimera_score}) is maximized.

\begin{figure} 
\includegraphics[width=\columnwidth]{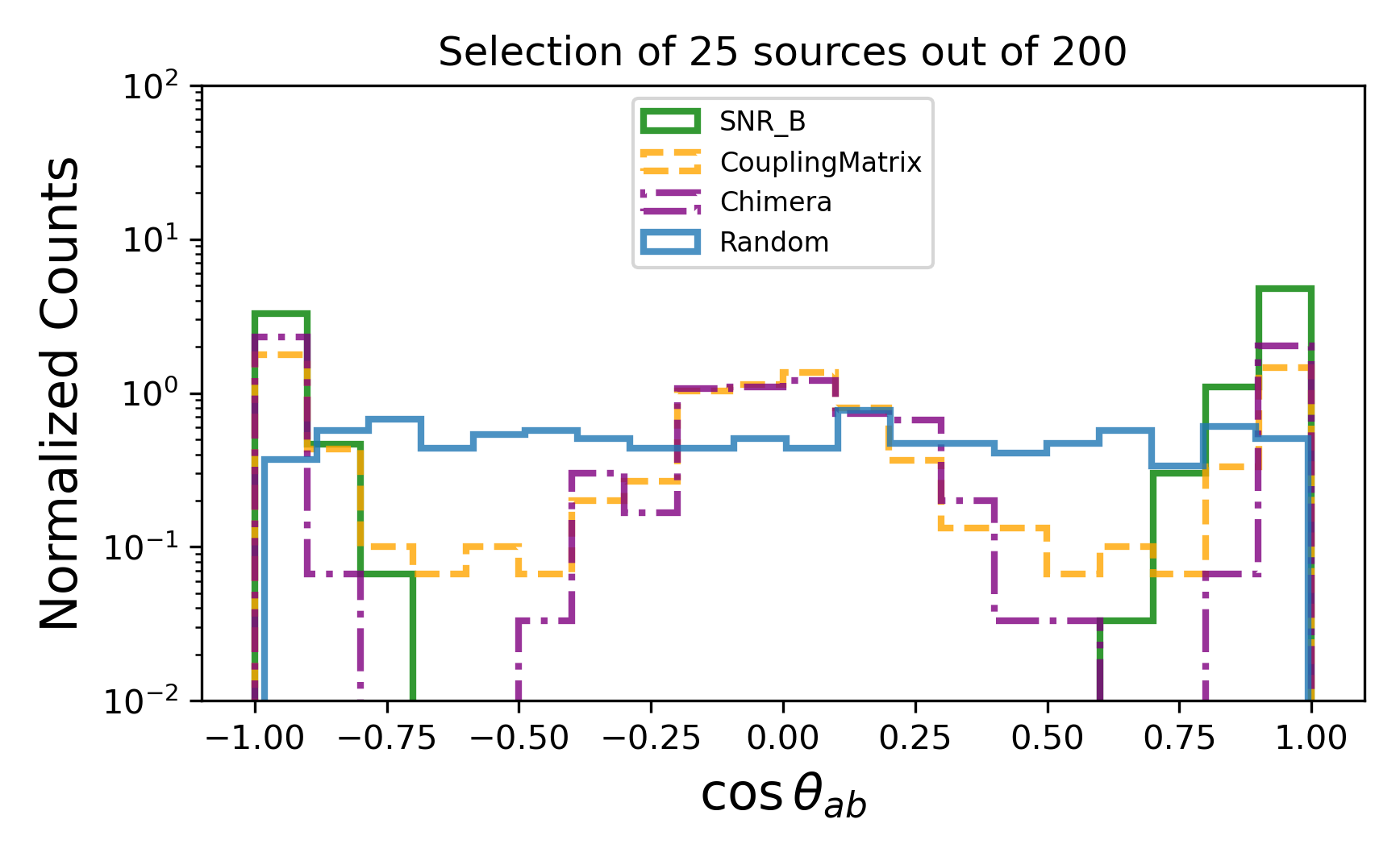} 
\caption{Distribution of angular separations {of 25 pulsars} selected with three selection methods, namely SNR$_\textrm{B}$-maximization, Coupling Matrix and Chimera. These methods have been applied to a dataset consisting of 200 pulsars with uniform sky distribution and equal noise properties. For reference, we also show a random selection of 25 pulsars.
} 
\label{fig:ang_distr}
\end{figure}

For reference, in Figure~\ref{fig:ang_distr} we show how the three different selection methods for GWB searches pick equal-noise pulsars on the sky. The full array is composed of 200 pulsars uniformly distributed over the sky and the number of selected pulsars is 25. {The first pulsar was randomly selected and the following ones were picked according to the different selection methods.
The SNR depends on $\Gamma_{ab}^2$ and so the SNR-maximization method tends to add pulsars where the HD correlation is largest, i.e., with $\theta_{ab}=0^\circ$ and $180^\circ$}. The region between -0.6 and 0.6 will be eventually filled as the number of selected pulsars increases\footnote{We included in the supplementary materials two animated figures that show how the SNR-maximization method squentially adds pulsars, see \url{animate_hist_HDvsNoise_loc_3d.gif} and \url{animate_hist_HDvsNoise.gif}.}. The Coupling Matrix and Chimera methods also {picked} pulsars at $\theta_{ab}=0^\circ$ and $180^\circ$, but the distribution of angular separations is broader and covers more values of $\theta_{ab}$. We find that {of the first 25 pulsars selected by} the Chimera method, {none of them are placed} around $\cos \theta _{ab}\approx -0.7$ and $\cos \theta _{ab}\approx 0.7$. This might be due to some interaction between SNR-maximization and Coupling Matrix selection. {Note that the pattern in Fig.~\ref{fig:ang_distr} could change if we were starting with two or more pulsars with different sky locations.}

%----------------------------------------------------------

%----------------------------------------------------------
\subsection{Continuous gravitational wave SNR maximization}
\label{subsec:continuous_snr}
%----------------------------------------------------------

Continuous gravitational waves are deterministic signals and their analysis has been treated separately from the stochastic GWB.
CGWs are included in the model as a periodic delay applied to the timing residuals $\vec{\delta t}$ while the effect of the GWB is included in the covariance matrix $\vb{C}$ of the likelihood. This fundamental difference between the two signals and their mathematical description calls for a different ranking method.

Here, we want to rank pulsars according to their response to a CGW signal. One way to proceed is to inject a large number of fake CGW signals with randomized parameters except for fixed frequency and amplitude \citep{babak2015}. Then, for each pulsar, the CGW signal-to-noise ratio is computed for each injection and averaged numerically. In this way, we have the average response of each individual pulsar in the array at a given frequency of the CGW signal. This averaging can also be done analytically, as shown in the following paragraph. Note that we refer to the signal-to-noise ratio of CGWs using the acronym SNR. However, we use the symbol $\rho$ to distinguish the SNR of CGWs from the previously defined SNRs.

In the likelihood of Eq.~(\ref{eq:likelihood}), the inclusion of a deterministic signal is performed by changing the timing residuals as $\delta \vec{t} \rightarrow \delta \vec{t} - \vec{s}(\vec{\theta})$ where $\vec{s}(\vec{\theta})$ is the signal template we aim to measure. In that case, the likelihood can be rewritten as:
\begin{equation}
    \ln \mathcal{L} = - \frac{1}{2} \qty[(\delta\vec{t} | \delta\vec{t}) + (\vec{s} | \vec{s}) - 2(\delta\vec{t} | \vec{s}) +  \Tr \ln{2 \pi \vb{C}} ] \, ,
\end{equation}
where we have introduced the noise weighted inner product $(\vec{x}|\vec{y}) = x^T \vb{C} ^{-1} y$.

We can now calculate this expression for the hypothesis of the presence of a CGW ($\mathcal{H}_1$) versus its absence ($\mathcal{H}_2$). The expectation value of the log-likelihood ratio becomes:
\begin{equation}
\begin{split}
    \langle \ln{\Lambda} \rangle_{\mathcal{H}_1} 
    = 
    \Bigg\langle \ln\qty(\frac{p(\delta \vec{t}|\vec{s})}{p(\delta \vec{t}|\vec{0})})  \Bigg\rangle_{\mathcal{H}_1} = \langle (\delta \vec{t}|\vec{s}) -  \frac{1}{2}(\vec{s}|\vec{s})\rangle_{\mathcal{H}_1} = \frac{1}{2}(\vec{s}|\vec{s}) \, ,
\end{split}
\end{equation}
where $\rho_\text{\tiny Opt} = \sqrt{(\vec{s}|\vec{s})}$ is the optimal SNR for the CGW source.

Since the source parameters are not known a priori, we average $\rho_\text{\tiny Opt}^2$ over gravitational wave polarization $\psi$, initial phase $\phi_0$, inclination $\iota$, and sky location $(\theta,\phi)$. To do so, we analytically compute the integral over the defined bounds of the CGW parameters:
\begin{equation}
\label{eq:averaged_cw_snr_bounds}
\rho^2 = \int_0 ^\pi \frac{d\psi}{\pi}
                                \int _0 ^{2\pi} \frac{d\phi_0}{2\pi}
                                \int _1 ^{-1} \frac{d\cos \iota}{2}
                                \int _1 ^{-1} \frac{d\cos \theta}{2}
                                \int _0 ^{2\pi} \frac{d\phi}{2\pi}
                                (\vec{s}|\vec{s}) \, .
\end{equation}
Using the formula for a CGW signal from a circular SMBHB, $\vec{s}(t, \Omega)$, as presented in \cite{Babak_2012}, the {Earth-term} SNR$^2$ averaged over CGW parameters takes this simple form:
\begin{equation}
    \rho^2(h, f)  = \frac{4}{15} \bigg(\frac{h}{2 \pi f}\bigg)^2\bigg[ \qty(\cos 2 \pi f t|\cos 2 \pi f t) + \qty(\sin 2 \pi f t|\sin 2 \pi f t) \bigg] \, ,
\end{equation}
with
\begin{equation}
    h = \frac{2\mathcal{M}^{5/3} (\pi f)^{2/3}}{d_L},
\end{equation}
where $f$ and $h$ are the gravitational wave frequency and amplitude, $\mathcal{M}$ is the chirp mass and $d_L$ is the luminosity distance. For pulsar $a$, we evaluate $ \rho^2_a $ at the TOAs $\vec{t}_a$. 
{
We consider an Earth-term only SNR for simplicity as the inclusion of the pulsar term is unlikely to make a significant difference to the ranking. In the absence of a chirp, the contribution of the pulsar term to the SNR$^2$ is equal to that of the Earth term, therefore leaving the relative contribution of different pulsars unchanged. When the system is chirping this is no longer true as different pulsar terms contribute at different frequencies. However, it is slightly misleading to include these in the ranking on an equal footing with the Earth terms, since matching the pulsar terms in the data is much harder and requires good knowledge of the pulsar distance. In addition, the resulting ranking would be dependent on the nature of the source in the data, as this determines the frequencies of each of the pulsar terms, which would not be known until after the analysis using the reduced set of pulsars had been completed.
} The correlated noises (e.g. intrinsic and dispersion measure noises) are taken into account in the covariance matrix $\vb{C}$ of the noise-weighted inner product of the cosine and sine terms. 

Common {(correlated)} processes were not included in our noise model, so the covariance matrix is block-diagonal. In this way, the likelihood can be factorized and SNR$^2$s can be computed independently for each pulsar.
{Common uncorrelated processes can be included without affecting the block diagonal form of the matrix, and this could be used as a proxy for the presence of a GWB background or other processes.} 
In practice, we should incorporate these common processes in the noise model, but this adds another level of complexity that is irrelevant for the goal of the selection procedure\footnote{Furthermore, detectable CGW signals must be louder than the GWB. Since the GWB is stronger at lower frequencies, CGW signals are more likely to be found at high frequencies.}. The ultimate goal is identification of the best pulsars for CGW detection, and therefore, only the intrinsic properties of the pulsars were considered.

We estimate the relative contribution of one pulsar to the total SNR of the array using the normalized SNR$^2$:
\begin{equation}
    \Bar{\rho}_a ^2 (f) = \frac{\rho_a ^2 (h,f)}{\sum_b \rho_b ^2 (h,f) },
\label{eq:norm_snr}
\end{equation}
Note that the amplitude $h$ cancels out in this expression and the CGW frequency $f$ remains the only parameter. Therefore we can fix $h$ to any value without affecting the ranking.

We construct the cumulative sum of the normalized SNR$^2$s of the pulsars ranked from best to worst.
We fix a threshold value for the SNR$^2$ cumulative sum above which pulsar contributions to the total SNR$^2$ are not considered significant. This value was chosen to be $0.95$. The process is illustrated in Figure~\ref{fig:cumul_plot} and in the animated Figure (\url{cgw_ranking.gif} included in the supplementary materials) for pulsars from the IPTA second data release \citep[DR2,][]{2019MNRAS.490.4666P}.

\begin{figure}
    \centering
    \includegraphics[width=0.5\textwidth]{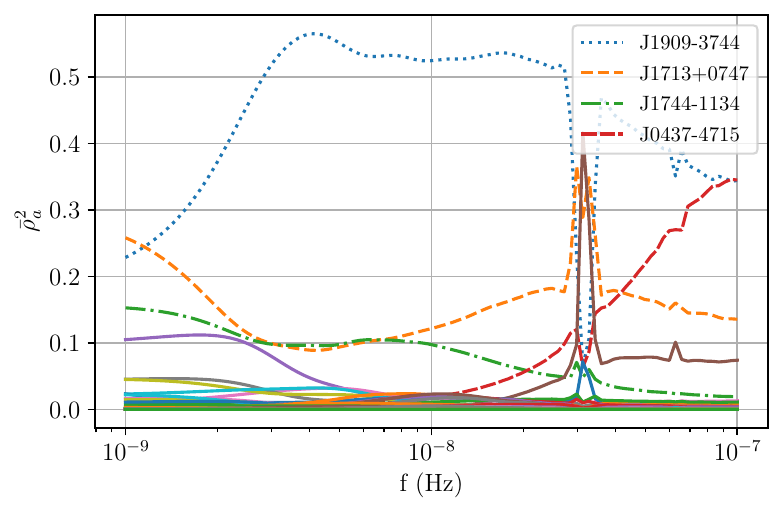}
    \caption{Normailzed $\Bar{\rho}_a ^2$ of the five best pulsars of the IPTA DR2, at different CGW frequencies. The glitches at the right of the plots are due to the one year and half-year peaks.}
    \label{fig:snr_freq}
\end{figure}

Due to the strong dependence of $\Bar{\rho}_a ^2 (f)$ on $f$, the resultant CGW pulsar ranking is also frequency dependent. This can be clearly seen from Figure~\ref{fig:snr_freq}. In our analysis, we use 100 log-spaced frequency bins between $10^{-9}$ and $10^{-7}$ Hz. Ranking lists were obtained separately for each frequency bin. In order to construct the final ranking catalog of best pulsars at a given frequency range, the lists at each frequency are merged together. This procedure ensures that we will gain at least, no matter the CGW frequency, 95\% of the total SNR$^2$ of the array. 

%%%%%%%%%%%%%%%%%%%%%%%%%%%%%%%%%%%%%%%%%%%%%%%%%%%%%%%%%%%%%%%%%%%%%
\begin{figure}
    \centering
    \includegraphics[width=0.5\textwidth]{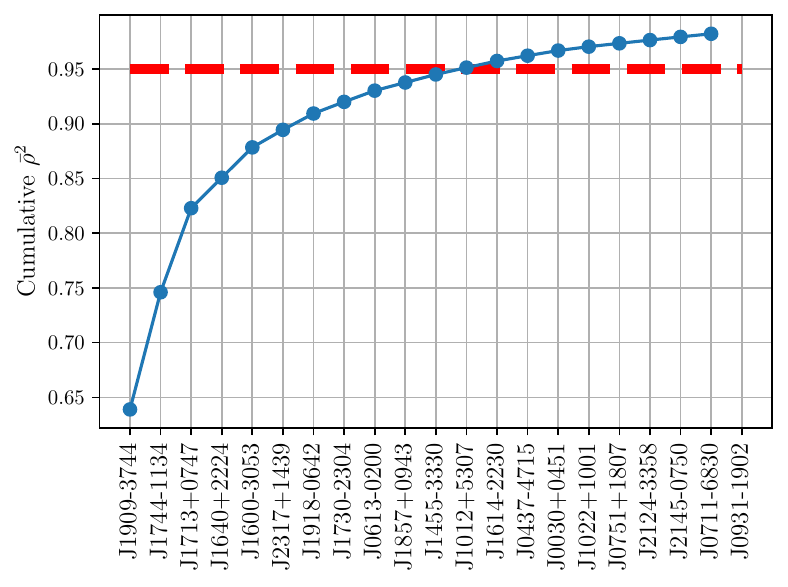}
    \caption{Cumulative $\Bar{\rho} ^2$ plot for the pulsars in the IPTA DR2 at CGW frequency of {5nHz}. The pulsars above the red dashed line contribute less than $5\%$ of the total SNR$^2$. This means only 12 pulsars out of {65} contribute on average to 95\% of the total SNR$^2$ of the array at {5nHz}. {Note that, while only the best 22 pulsars are shown in the figure, the normalized total SNR has been evaluated using all {65} pulsars in the array.}
    }
    \label{fig:cumul_plot}
\end{figure}
%%%%%%%%%%%%%%%%%%%%%%%%%%%%%%%%%%%%%%%%%%%%%%%%%%%%%%%%%%%%%%%%%%%%%

%%%%%%%%%%%%%%%%%%%%%%%%%%%%%%%%%%%%%%%%%%%%%%%%%%%%%%%%%%%%%%%%%%%%%
%%%%%%%%%%%%%%%%%%%%%%%%%%%%%%%%%%%%%%%%%%%%%%%%%%%%%%%%%%%%%%%%%%%%%

%----------------------------------------------------------
\section{Results}
\label{sec:results}
%----------------------------------------------------------
We create mock PTA datasets with increasing complexity in the noise models and test the performance of the selection methods. 
The PTA datasets are simulated using \textsc{LIBSTEMPO}\footnote{\url{https://github.com/vallis/libstempo}} and analysed using \textsc{ENTERPRISE} \citep{enterprise} giving the marginalized likelihood. Bayes factors are computed using \textsc{DYNESTY} \citep{dynesty}. 

%-------------------------------------------------------------
\subsection{Testing the selection methods for GWB searches}
\label{subsec:test_methods}
%-----------------------------------------------------------
In this section we investigate the performance of the three ranking methods that target GWB searches (Sec.~tion \ref{subsec:ranking_stochastic}). We consider a simplified framework, in which the pulsar noise is white noise only, and there is an injected GWB with amplitude $A_\textrm{GWB}=3\times10^{-15}$ and slope $\gamma=13/3$, consistent with findings from the EPTA analysis \citep{chen2021}. 
We pick pulsars one by one using the SNR$_\text{B}$-maximization, the Coupling Matrix method {(with weights $w \sim \textrm{SNR}_\text{A}$)}, and the Chimera method, and we investigate the performance of these procedures by calculating the log-Bayes factor {($\ln $BF natural logarithm)} of the following hypothesis tests:
\begin{itemize}
    \item HD vs CURN: Hellings \& Downs correlation versus a common uncorrelated red noise process;
    \item HD vs CURN $+$ MN $+$ DN: Hellings \& Downs correlation versus a combination of common uncorrelate red process, monopolar noise (MN) and dipolar noise (DN).
\end{itemize}
Since a detectable GWB signal is injected, we expect the log-Bayes factor to always increase in the limit of a high {number of pulsars} $N$. Of particular importance, however, are the dynamics of growth of the log-Bayes factor with respect to a random selection. {A further comparison of these selection methods against a lowest-RMS selection procedure is presented in Appendix \ref{subsec:rms_selection}}

Note that the white noise parameters are kept fixed, and only the amplitudes and slopes of the common red noise processes are varied. In the next sections we present the evolution of the log-Bayes factor obtained with the $N$ pulsars selected with the aforementioned methods. We anticipate that the performance of the selection methods strongly depends on the specifics of the dataset considered. Therefore, we tested our ranking methods with three different simulated datasets.

%^^^^^^^^^^^^^^^^^^^^^^^^^^^^^^^^^%^^^^^^^^^^^^^^^^^^^^^^^^^^^^^^^^^
\subsubsection{Galaxy-distributed dataset}
\label{subsubsec:galaxy}
\begin{figure*}
\text{\large Galaxy-distributed dataset}\par\medskip
\begin{subfigure}[t]{.49\textwidth}
\centering
\includegraphics[width=1.1\textwidth]{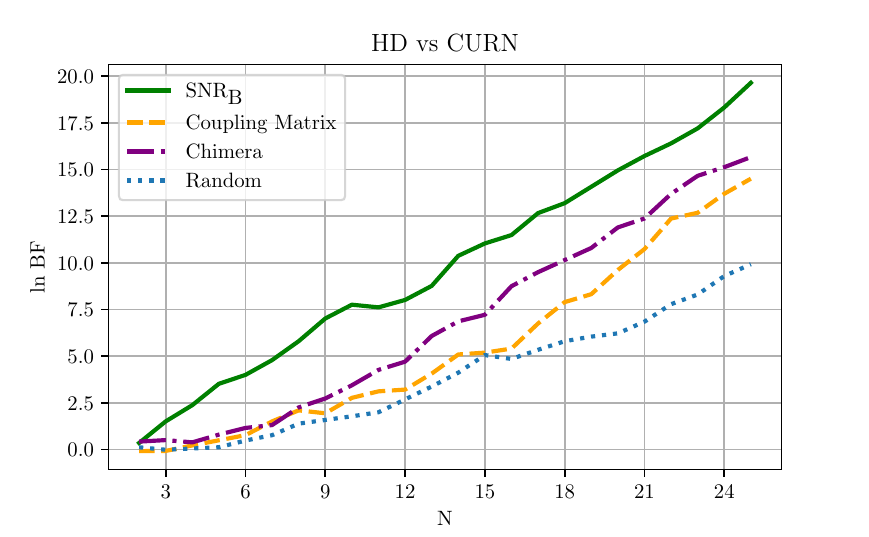}
\end{subfigure}
\begin{subfigure}[t]{.49\textwidth}
\centering
\includegraphics[width=1.1\textwidth]{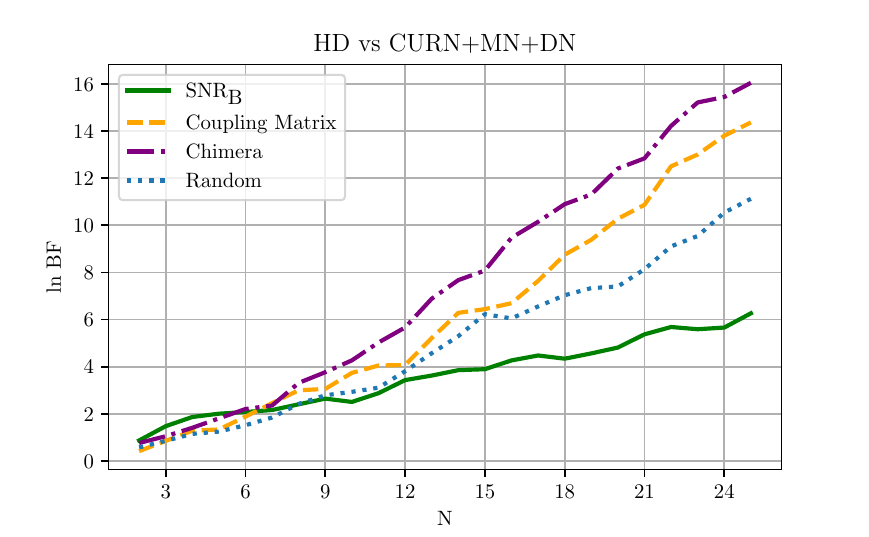}
\end{subfigure}
\caption{
Log-Bayes factor as a function of the number of chosen pulsars by each of the selection methods (shown in different colors) for the Galaxy-distributed dataset and for different hypothesis tests: HD vs CURN (left), and HD vs CURN+MN+DN (right). The 200 simulated pulsars have the same noise properties and galaxy-distributed sky locations. {The first pulsar is selected at random 20 times and the shown log-Bayes factors are the average over these 20 realizations. For 25 selected pulsars the mean and standard deviation values are: SNR$_\text{B}$: 20$\pm 6$, Coupling Matrix: 15$\pm 7$, Chimera: 16$\pm 5$, Random: 10$\pm 4$ (HD vs CURN hypothesis test (left)); SNR$_\text{B}$: 6$\pm 2$, Coupling Matrix: 14$\pm 7$, Chimera: 16$\pm 5$, Random: 11$\pm 4$ (HD vs CURN+MN+DN hypothesis test (right)). The log-Bayes factors of the whole array for one realization are 198 and 194 for HD vs CURN (left), and HD vs CURN+MN+DN (right), respectively.}
}
\label{fig:galaxy_dataset}
\end{figure*}

%^^^^^^^^^^^^^^^^^^^^^^^^^^^^^^^^^%^^^^^^^^^^^^^^^^^^^^^^^^^^^^^^^^^
We created an array of 200 pulsars with equal RMS of 100 ns with galaxy distribution on the sky. The sky coordinates were drawn randomly from the available values of known pulsars in the \texttt{psrcat} catalogue \citep{2004IAUS..218..139H}. The total timespan of the dataset is {10} years with a sampling rate of 28 days. A dataset consisting of all equal pulsars with a dense sky coverage serves to demonstrate how each selection method performs under idealised conditions. In Figure~\ref{fig:galaxy_dataset} we show the log-Bayes factor computed using the pulsars selected by the different ranking methods when applied to the Galaxy-distributed dataset for the hypothesis tests: HD vs CURN, and HD vs CURN $+$ MN $+$ DN. 
%%%%%%%%%%%%%%%%%%%%%%%%%%%%%%%%%%%%%%%%%%%%%%%%%%%%%%%%%%%%%%%%%%%%%
%%%%%%%%%%%%%%%%%%%%%%%%%%%%%%%%%%%%%%%%%%%%%%%%%%%%%%%%%%%%%%%%%%%%%
The very first pulsar in the array was selected at random 20 times, so that the log-Bayes factor shown in Figure~\ref{fig:galaxy_dataset} is an average over these realizations. This procedure was done in order to ensure that our results are independent of the initial pulsar choice. For reference, we also show the log-Bayes factor obtained with a random selection of pulsars.

The left panel of Figure~\ref{fig:galaxy_dataset} demonstrates that the Coupling Matrix method (dashed yellow line) performs similarly to the random selection (dotted blue line) for the HD vs CURN hypothesis test,  with slightly better performance after $\sim 15$ pulsars are included in the array. Both the SNR-maximization (solid green line) and Chimera method (purple dash-dotted line) outperform the other two types of selection. For the SNR maximisation method the log-Bayes factor increases with the number of pulsars in the array like $\sim 0.8N$, which results in almost double log-Bayes factor for $N=25$ than the one obtained using random selection.
These results are expected, since the SNR-maximization is designed to maximize the confidence of detecting the HD correlation versus a CURN process. 

The hypothesis test HD vs CURN $+$ MN $+$ DN is proposed to demonstrate the benefits of the Coupling Matrix, as the method is designed to disentangle the HD correlation from other types of common correlated noises. The right panel of Figure~\ref{fig:galaxy_dataset} confirms these expectations. We see that, in this context, the Coupling Matrix and Chimera methods provide a log-Bayes factor for $N=25$ pulsars which is $1.4$ and $1.6$ times larger than a random selection, respectively. The scaling of the log-Bayes factor for the Chimera selection is $\sim 0.8N$, while the SNR selection scales only as $\sim 0.2N$. The SNR-maximization is severely suboptimal for this test, as it tends to pick pulsars at locations where the HD overlap reduction function is the largest, i.e., at $180^\circ$ and $0^\circ$, making it harder to discern HD from other types of correlation. A random selection of pulsars provides a more distributed sky coverage which improves the situation in this regard.

The slightly improved performance of the Chimera method in comparison to the Coupling Matrix formalism is due to the fact that it accounts for both the optimal sky coverage and total gain in SNR. These results confirms that both of these components are essential for PTA optimization and cannot be ignored. One can conclude that the inclusion of the SNR-maximiztion in the Chimera method is of special relevance in the case of non-equal pulsar arrays. The latter point is even more evident in one of the following subsection, where we consider a simplified EPTA dataset.

%^^^^^^^^^^^^^^^^^^^^^^^^^^^^^^^^^%^^^^^^^^^^^^^^^^^^^^^^^^^^^^^^^^^
\subsubsection{Mock MeerTime dataset}
\label{subsubsec:meertime}
%^^^^^^^^^^^^^^^^^^^^^^^^^^^^^^^^^%^^^^^^^^^^^^^^^^^^^^^^^^^^^^^^^^^
We now consider a PTA dataset which resembles the properties of the recently published 5-year MeerTime Large Survey \citep{2022arXiv220404115S}. This survey is expected to significantly increase the sensitivity of current PTAs in the very near future. Using this as motivation, we created a mock MeerTime dataset consisting of 189 pulsars with sky positions taken from the survey. Observations were performed every 28 days on a baseline of {10} years. { The white noise RMS is set to the median TOA uncertainties delivered by MeerTime, in which each observation epoch of each source consisted of 256 seconds of integration time with the MeerKat radio telescope}. The dataset provides an insight on how the pulsar selection performs with a large dataset composed of non-equal pulsars with realistic sky positions.
%%%%%%%%%%%%%%%%%%%%%%%%%%%%%%%%%%%%%%%%%%%%%%%%%%%%%%%%%%%%%%%%%%%%%
\begin{figure*}
\centering
\text{\large Mock MeerTime dataset}\par\medskip
\begin{subfigure}[t]{.49\textwidth}
\centering
\includegraphics[width=1.1\textwidth]{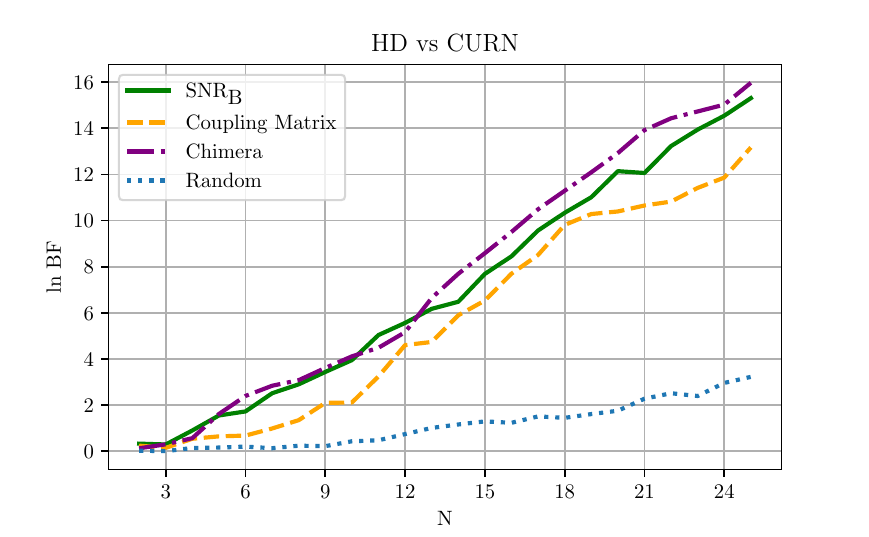}
\end{subfigure}
\begin{subfigure}[t]{.49\textwidth}
\centering
\includegraphics[width=1.1\textwidth]{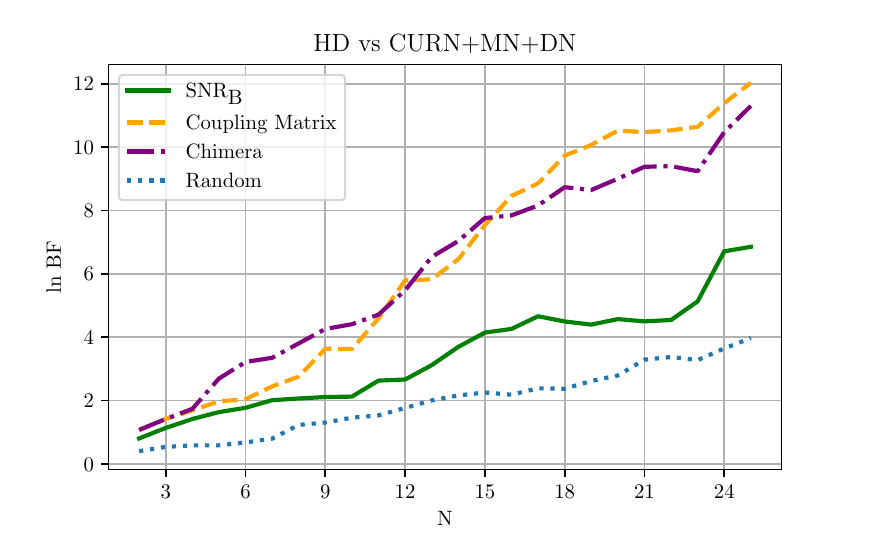}
\end{subfigure}
\caption{
Log-Bayes factor as a function of the number of chosen pulsars by each of the selection methods (shown in different colors) for the mock MeerTime dataset and for different hypothesis tests: HD vs CURN (left), and HD vs CURN+MN+DN (right).
{The shown log-Bayes factors represent the average over 20 different noise realizations. For 25 selected pulsars the mean and standard deviation values are: SNR$_\text{B}$: 15$\pm 10$, Coupling Matrix: 13$\pm 8$, Chimera: 16$\pm 10$, Random: 3$\pm 2$ (HD vs CURN hypothesis test (left)); SNR$_\text{B}$: 7$\pm 3$, Coupling Matrix: 12$\pm 6$, Chimera: 11$\pm 6$, Random: 4$\pm 2$ (HD vs CURN+MN+DN hypothesis test (right)). The log-Bayes factors of the whole array are 57$\pm$21 and 47$\pm$16 for HD vs CURN (left), and HD vs CURN+MN+DN (right), respectively.}
}
\label{fig:meerkat_dataset}
\end{figure*}
%%%%%%%%%%%%%%%%%%%%%%%%%%%%%%%%%%%%%%%%%%%%%%%%%%%%%%%%%%%%%%%%%%%%%
% %%%%%%%%%%%%%%%%%%%%%%%%%%%%%%%%%%%%%%%%%%%%%%%%%%%%%%%%%%%%%%%%%%%%%
\begin{figure*}
\centering
\text{\large EPTA-simplified dataset}\par\medskip
\begin{subfigure}[t]{.49\textwidth}
\centering
\includegraphics[width=1.1\textwidth]{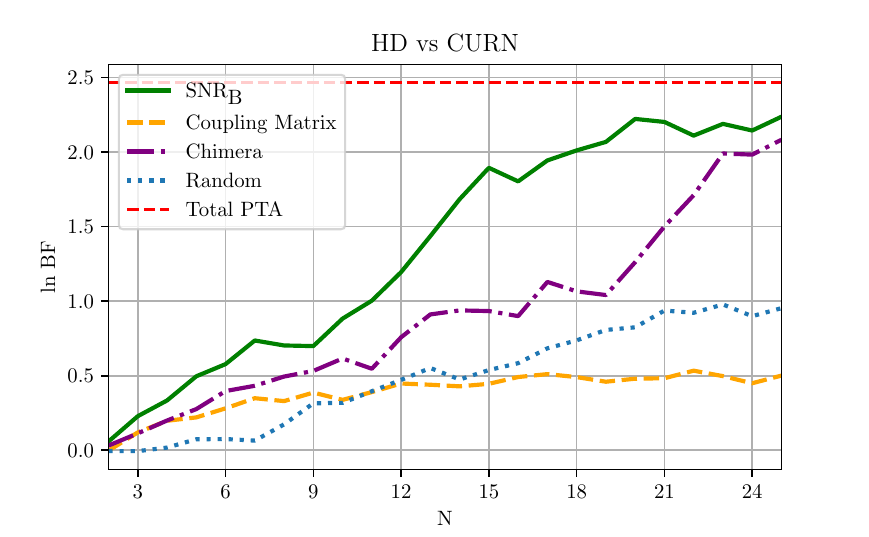}
\end{subfigure}
\begin{subfigure}[t]{.49\textwidth}
\centering
\includegraphics[width=1.1\textwidth]{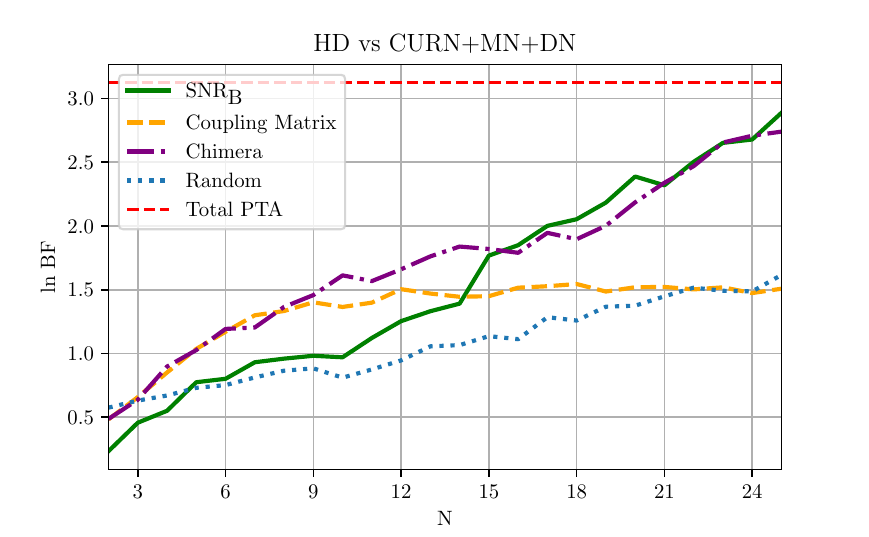}
\end{subfigure}
\caption{
Log-Bayes factor as a function of the number of chosen pulsars by each of the selection methods (shown in different colors) for the EPTA-simplified dataset and for different hypothesis tests: HD vs CURN (left), and HD vs CURN+MN+DN (right). {The shown log-Bayes factors represent the average over 20 different noise realizations. For 25 selected pulsars the mean and standard deviation values are: SNR$_\text{B}$: 2.2$\pm 1.9$, Coupling Matrix: 0.5$\pm 1.1$, Chimera: 2.1$\pm 1.9$, Random: 1.0$\pm 1.2$ (HD vs CURN hypothesis test (left)); SNR$_\text{B}$: 2.9$\pm 1.8$, Coupling Matrix: 1.5$\pm 1.0$, Chimera: 2.7$\pm 1.9$, Random: 1.6$\pm 1.4$ (HD vs CURN+MN+DN hypothesis test (right)). The red-dashed line shows the log-Bayes factor of the full dataset (N=40): 2.5$\pm 2.3$ for HD vs CURN and 3.1$\pm 2.2$ for HD vs CURN+MN+DN.} 
}
\label{fig:simpli_epta_dataset}
\end{figure*}
% %%%%%%%%%%%%%%%%%%%%%%%%%%%%%%%%%%%%%%%%%%%%%%%%%%%%%%%%%%%%%%%%%%%%%

We generate 20 noise realizations of this dataset and show the averaged log-Bayes factor in Figure~\ref{fig:meerkat_dataset}. The first pulsar in the ranking is fixed to the one with the smallest RMS.

The left panel of Figure~\ref{fig:meerkat_dataset} shows the ranking for the HD vs CURN test, and it confirms that the Chimera method and the SNR-maximization are optimal in this case. Even though the pulsars selected with the Coupling Matrix method provide a log-Bayes factor smaller than the other methods, it still gives an evidence which is approximately 3 times larger in comparison to random selection for $N=25$.

The evolution of the log-Bayes factor for the hypothesis test HD vs CURN $+$ MN $+$ DN is shown in the right panel of Figure~\ref{fig:meerkat_dataset}. The Coupling Matrix and Chimera selections increase the log-Bayes factor up to $\log_{10}$BF~$\approx12$. Differently from the ``galaxy-distributed'' dataset, the SNR-maximization performs slightly better than the random selection, although still worse than the Coupling Matrix and Chimera methods. Up to the first 18 pulsars, the Chimera method provides a stronger support for HD vs CURN $+$ MN $+$ DN than the Coupling Matrix, reaching similar levels for larger number of pulsars.

%^^^^^^^^^^^^^^^^^^^^^^^^^^^^^^^^^%^^^^^^^^^^^^^^^^^^^^^^^^^^^^^^^^^
\subsubsection{EPTA-simplified dataset}
\label{subsubsec:mock_epta}
%^^^^^^^^^^^^^^^^^^^^^^^^^^^^^^^^^%^^^^^^^^^^^^^^^^^^^^^^^^^^^^^^^^^
We construct an EPTA-simplified dataset, which consists of 40 pulsars with RMS and sky location of the latest EPTA dataset \citep{2016MNRAS.458.3341D, chen2021}. The total timespan is fixed to {10} years with observations being performed every 28 days. In order to reduce required computational resources, only white noise was taken into account, ignoring the red intrinsic and interstellar medium noise contributions. Despite the significant simplification, this dataset serves to imitate a realistic PTA setup with a modest number of pulsars and representative pulsar sensitivities, which has been principally used for GW searches to date. We have simulated 20 statistically equivalent noise realizations. The averaged log-Bayes factor are shown in Figure~\ref{fig:simpli_epta_dataset}. As in the case of the mock MeerTime dataset, the first initial pulsar is chosen to be the one with the smallest RMS.

It can be seen from both panels of
Figure~\ref{fig:simpli_epta_dataset}, that the restricted dataset of 25 pulsars chosen by the Chimera or SNR-maximization methods on average reaches higher log-Bayes factors than those selected randomly or using the  Coupling Matrix formalism. Moreover, Figure~\ref{fig:simpli_epta_dataset}
shows that by using only 25 of pulsars picked by one of the two former methods, we account for $\approx90\%$ of the sensitivity of the whole array. The Coupling Matrix approach, on the other hand, falls behind, even for the HD vs CURN+MN+DN hypothesis test. These results clearly demonstrate that pulsar quality is as important as optimal sky location, when disentangling different types of correlations. The Coupling Matrix is not aimed at maximizing the SNR, therefore it can not be used as a selection method on its own, as some of the highly sensitive sources could be discarded. The best results are obtained when the optimal sky location and gain in SNR are finely balanced. Therefore, ``good'' pulsars must be picked at proper sky locations, which is the main idea behind the Chimera method. In other words, neither low-sensitivity sources selected at proper angular distances, nor high-SNR sources with poorly chosen coordinates, e.g. clustered at a specific location on the sky, can provide an adequate improvement in performance. The former case is the Coupling Matrix selection for the EPTA-simplified dataset (yellow dashed line in the left panel of Figure~\ref{fig:simpli_epta_dataset}), while the latter corresponds to SNR-maximization for the MeerTime dataset (solid green line in the right panel of Figure~\ref{fig:meerkat_dataset}).

We want to remark that the Chimera implementation we offer in this paper is not the ultimate solution. Alternative ways to address this issue are proposed in Appendix~\ref{subsec:cm_wgt}. {Furthermore, as demonstrated in Appendix~\ref{subsec:rms_selection}, simpler ranking criteria might perform better than the Chimera method for some datasets.} More thorough investigations are left for future works.

%-----------------------------------------------------------
\subsection{Optimizing the search for a GWB in a {realistic} EPTA dataset}
\label{subsec:optimize_epta}
%-----------------------------------------------------------
To speed-up the assembly of the new dataset and to improve computational efficiency of the analysis, the EPTA collaboration decided to select a subsample of pulsars timed by its radio facilities. In this context, it is of paramount importance to wisely pick the pulsars to be included. Therefore, we create another simulated array to address this problem. We consider a dataset similar to the one of Sec.~\ref{subsubsec:mock_epta}{, i.e. 40 pulsars with RMS, timespan, and sky locations of the EPTA dataset}, but more realistic in the sense that we include the intrinsic red-noise properties of the preliminary EPTA dataset\footnote{For simplicity we adopt the best fit estimates as representative values from the EPTA constraints on the red noise parameters {and set the time interval between observations to be 14 days.}} \citep{chen2021, lentati2015}.

For simplicity, we focus on ranking the best pulsars to distinguish an HD correlation (hypothesis $\mathcal{H}_1$) from a CURN process (hypothesis $\mathcal{H}_2$) and we study how this can be affected by possible noise realizations.
% Although the sources should be selected also to disentangle different correlations, the EPTA dataset has a limited number of pulsars (N=40) and the SNR-maximization was able to disentangle different correlations for the mock EPTA dataset.
As shown in the previous sections, SNR-maximization and the Chimera method should be a good selection proxy for this hypothesis test. Since the SNR-maximization method is constructed to target this hypothesis and it has been shown to perform as well as the Chimera method, we will only use this method for this study. The first six pulsars are fixed to those which constitute the preliminary combination of \cite{chen2021}: J1909-3744, J1713+0747, J1744-1134, J0613-0200, J1600-3053, J1012+5307.

\begin{figure}
    \centering
    \includegraphics[width=0.5\textwidth]{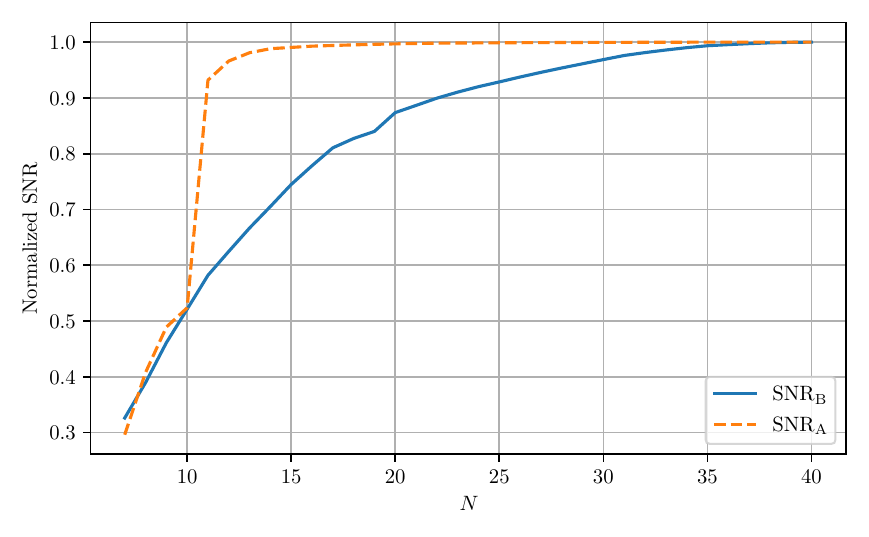}
    \caption{Normalized SNR evolution as a function of the number of selected pulsars $N$ with the SNR maximization method of statistic B and A. The SNR is normalized to the total SNR of the dataset and the initial pulsar subset is composed of the $6$ initial pulsars of the EPTA analysis  \citep{chen2021}.
    }
    \label{fig:snr_evolution}
\end{figure}

Firstly, we estimate the number of sources that to be added to the preliminary combination in order to achieve a reasonable detection confidence. For this, we apply the SNR maximization selection using the injected GWB parameters, and iteratively add the pulsars which increase the SNR the most. Results are shown in Figure~\ref{fig:snr_evolution}. SNR$_\text{A}$ tends to saturate more quickly than SNR$_\text{B}$. This is because the latter is suppressed by the term $S(f)$ in the denominator of Eq.~(\ref{eq:signal-to-noise-B}).
We find that with $N=25$ pulsars we reach $94\%$ of the total  SNR$_\text{B}$. Therefore, adding $19$ SNR-maximization selected pulsars to the starting six sources increases the SNR from $30\%$ to $94\%$ of the total SNR of the array.
%%%%%%%%%%%%%%%%%%%%%%%%%%%%%%%%%%%%%%%%%%%%%%%%%%%%%%%%%%%%%%%%%%%%%
%%%%%%%%%%%%%%%%%%%%%%%%%%%%%%%%%%%%%%%%%%%%%%%%%%%%%%%%%%%%%%%%%%%%%
\begin{figure}
    \centering
    \includegraphics[width=0.5\textwidth]{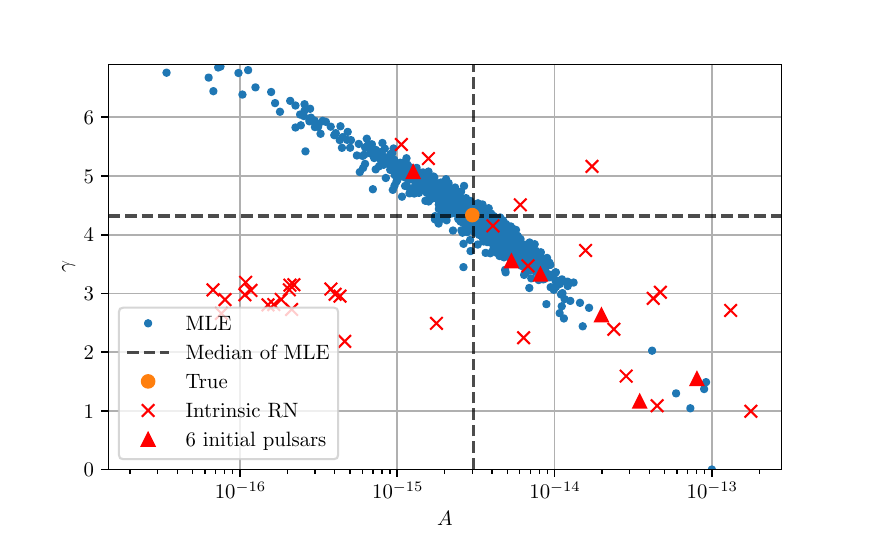}
    \caption{Maximum likelihood estimation of the amplitude $A$ and slope $\gamma$ of the stochastic gravitational-wave background using the first 6 pulsars of the EPTA mock dataset (the red triangles show the respective intrinsic red noise properties). The blue dots show the estimated values of $A$ and $\gamma$ per noise realization, and the dashed lines indicate the median distribution value. The orange dot shows the true injected value, whereas the red crosses show the values of the intrinsic red noises injected in the remaining pulsars.}
    \label{fig:amp_gam_MLE}
\end{figure}
%%%%%%%%%%%%%%%%%%%%%%%%%%%%%%%%%%%%%%%%%%%%%%%%%%%%%%%%%%%%%%%%%%%%%

Next, we want investigate whether the selection procedure is strongly affected by the choice of GWB parameters.
To this end, we simulate the EPTA mock dataset $1000$ times with the same injection parameters, and find the Maximum Likelihood Estimator using only the first six pulsars (preliminary dataset) and assuming an HD correlation only. The intrinsic red and white noise parameters were fixed to the true values. The results are shown in Figure~\ref{fig:amp_gam_MLE}. Different noise realizations lead the MLE values (blue dots) to be shifted from the true parameters (orange dot). It can be clearly seen that the distribution of MLEs lies along the line over which the six initial pulsars are located (red triangles), and its median (dashed black lines) is consistent with the injected true parameters. For reference, we show the adopted intrinsic red noise parameters of the other pulsars in the simulated datasets as red crosses.

\begin{figure}
    \centering
    \includegraphics[width=0.5\textwidth]{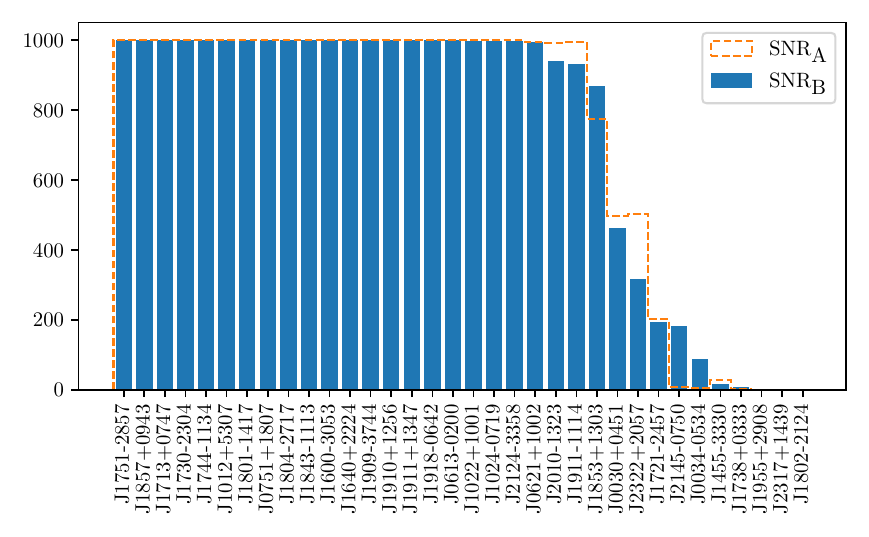}
    \caption{Histogram of the 25 pulsars selected with the SNR$_\text{B}$ (blue) and SNR$_\text{A}$ (orange) maximization over 1000 noise realizations.}
    \label{fig:snr_selection_histogram}
\end{figure}

We now use each of the MLEs of Figure~\ref{fig:amp_gam_MLE} as a new set of GWB parameters and run the SNR ranking procedure. The histogram of the best 25 selected pulsars is shown in Figure~\ref{fig:snr_selection_histogram}. Since the GWB parameters are different at every realization, the subset of selected pulsars slightly changes. As expected, the histogram for the SNR$_\text{B}$ selection has larger tails since different GWB parameters affect both the denominator and numerator of the Eq.~\ref{eq:signal-to-noise-B}. Instead, the SNR$_\text{A}$ is affected only by the variation in the GWB slope $\gamma$. 
{Both SNR$_\text{A}$ and SNR$_\text{B}$ selections exclude 15 pulsars in each realization. This selection reduces the total number of TOAs to analyze from 18584 to 12191 (in median). Therefore, the SNR ranking procedure excludes $6393/18584\approx35\%$ of the TOAs of the full dataset by excluding 15 out of 40 pulsars.}
As shown in Figure~\ref{fig:snr_selection_histogram}, both methods pick the same 20 pulsars in majority of the cases.
In practice, we could find the best pulsars by performing the selection process with the GWB and intrinsic red noise parameters taken from posterior chains of the previous data release. However, such an analysis is beyond the scope of this work.
%%%%%%%%%%%%%%%%%%%%%%%%%%%%%%%%%%%%%%%%%%%%%%%%%%%%%%%%%%%%%%%%%%%%%
%%%%%%%%%%%%%%%%%%%%%%%%%%%%%%%%%%%%%%%%%%%%%%%%%%%%%%%%%%%%%%%%%%%%%

\begin{figure}
    \centering
    \includegraphics[width=0.5\textwidth]{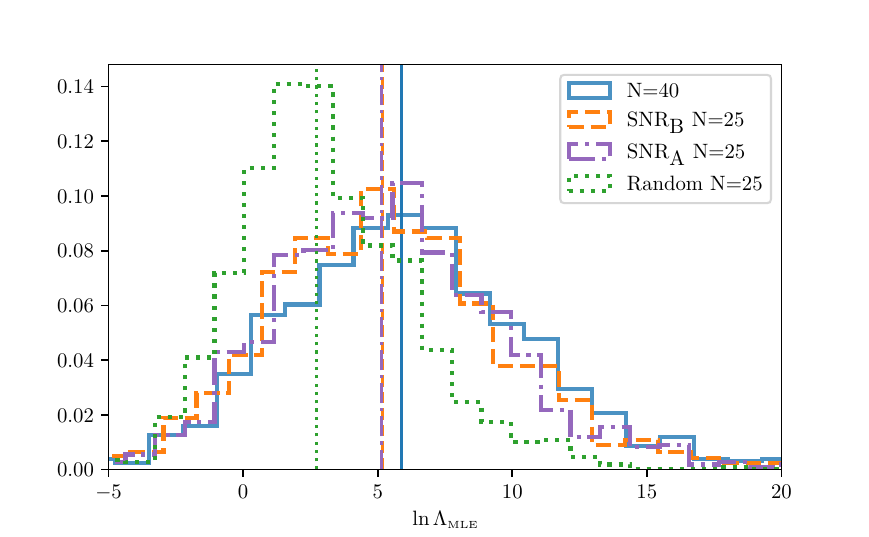}
    \caption{Distribution of log-likelihood ratios obtained with the full dataset $N=40$ (solid blue) and with 25 pulsars selected with SNR$_\text{B}$ (dashed orange) and SNR$_\text{A}$-maximization (dash-dotted purple) for 1000 noise realizations. For each noise realization we also randomly select 25 pulsars and calculate the log-likelihood ratio of this distribution. The distribution of these log-likelihoods is also shown as a green dotted histogram. The medians of the distributions are shown as vertical lines and are $5.88$ for $N=40$, $5.17$ for ${\rm SNR}_{\rm B} \, N=25$, $5.14$ for ${\rm SNR}_{\rm A} \, N=25$, and $2.73$ for Random $N=25$. The log-likelihood ratios have been all evaluated at the maximum likelihood value. 
    }
    \label{fig:lnL_comparison}
\end{figure}

We now demonstrate that the SNR-maximization selection method performs better than a random selection, and it provides evidence comparable to the full dataset. For each of the 1000 noise realizations, we select 25 pulsars in three ways: using the SNR-maximization methods (SNR$_\text{B}$ and SNR$_\text{A}$) as done in Figure~\ref{fig:snr_selection_histogram}, and randomly.
We compute the log-likelihood ratios obtained with the three different pulsar subsets and with the full dataset and we show the results in  Figure~\ref{fig:lnL_comparison}. These distributions are evaluated at maximum-likelihood estimates of the parameters (amplitudes and slopes of the GWB). Based on the median values of the distributions, one finds that the optimally selected datasets provide a factor of $1.84-1.90$ stronger evidence with respect to the random selection.
%%%%%%%%%%%%%%%%%%%%%%%%%%%%%%%%%%%%%%%%%%%%%%%%%%%%%%%%%%%%%%%%%%%%%
%%%%%%%%%%%%%%%%%%%%%%%%%%%%%%%%%%%%%%%%%%%%%%%%%%%%%%%%%%%%%%%%%%%%%
Furthermore, we find that the log-likelihood ratio for the 25 optimally selected dataset is in median $\sim 0.89$ times the one obtained from the full array. The distributions of log-likelihood ratios evaluated at the true parameters do not significantly differ from those shown in Figure~\ref{fig:lnL_comparison}. Therefore, the search over the GWB parameters with the MLE is not affecting the distribution of log-likeliood ratios.

These results demonstrate that the SNR-maximization selection method is a good proxy for choosing pulsars and it is robust against noise realizations. Furthermore, we have demonstrated that the log-likelihood ratio obtained with a subset of 25 pulsars is comparable to the one from the full array.

\begin{figure}
    \centering
    \includegraphics[width=0.5\textwidth]{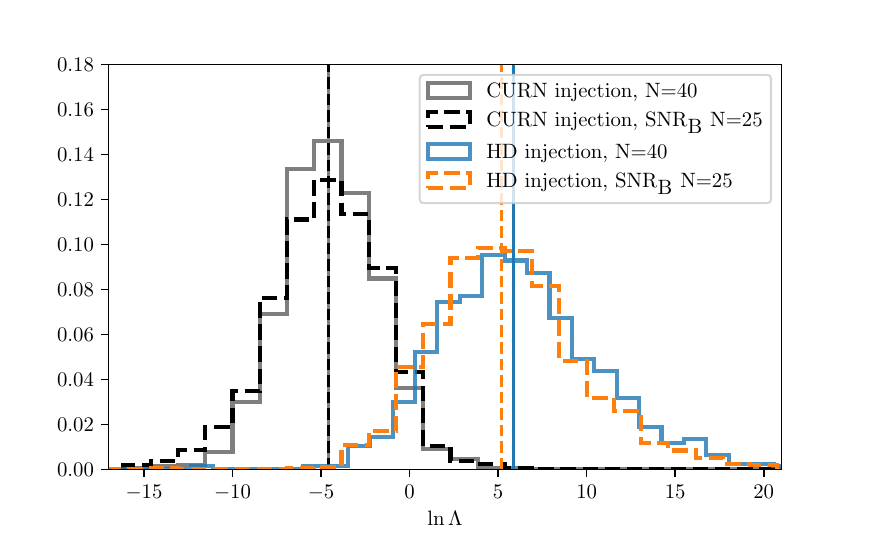}
    \caption{Distribution of log-likelihood ratios $\ln \Lambda$ for the hypothesis test of the HD correlation versus Common uncorrelated Red Noise process over many noise realizations and different injections. The dashed lines show the distribution when the log-likelihood is computed using the 25 pulsars selected with the SNR$_\text{B}$ maximization, whereas the solid lines when all 40 pulsars are used. {The median values of the distributions for the CURN injection are -4.56 and -4.60 for $N=40$ and SNR$_\text{B}$ $N=25$, respectively, whereas for the HD injection these are 5.87 and 5.17 for $N=40$ and SNR$_\text{B}$ $N=25$, respectively}. The log-likelihood ratios have all been evaluated at the true injected parameters. 
    }
    \label{fig:lnL_alt_hypotheis}
\end{figure}

Now, we establish the significance achieved by the optimally selected pulsars. To this purpose, we simulate two sets of realistic EPTA datasets: with an injected CURN process; and with an injected HD correlated process. The two injected common processes are characterized by the same amplitudes and slopes. We show in Figure~\ref{fig:lnL_alt_hypotheis} the log-likelihood ratios obtained using the full dataset (N=40) and the 25 SNR$_{\rm B}$ selected pulsars for the HD and CURN injection subsets. The median of the log-likelihood ratios of the best 25 pulsars for the HD injection (orange dashed-line histogram) corresponds to a p-value of $\approx 2\times 10^{-3}$ with respect to the CURN log-likelihood ratio distribution (black dashed-line histogram). 
The log-likelihood ratio distributions for the full array (N = 40) are shown in Fig.~\ref{fig:lnL_alt_hypotheis} as solid-line histograms for the CURN (gray) and HD injection (blue), respectively. Since the median of the latter distribution (HD) is above all the log-likelihood ratios obtained with the CURN injection with N=40 pulsars, we estimate the respective p-value as smaller than one over the number of noise realizations/samples, i.e. $\lesssim 10^{-3}$.
We caution the reader that the aforementioned p-values are only approximate. In fact, to resolve the tails of the CURN log-likelihood distribution, we would need to run our analysis for a larger number of noise realizations. Nevertheless, these results demonstrate that the selection of pulsars does not significantly affect the statistical significance of the hypothesis test.

We showed that the SNR-maximization selection method is a good proxy for ranking pulsars and it allows to reach detection confidence comparable to the full array. However, it is important to remark that these results are obviously dependent on the specific pulsars' sky localizations and noise properties and on the tested hypothesis (here HD vs CURN). We expect this ranking method to be well suited also for other PTA datasets where the pulsars have very different noise properties.

{We remark that similar results can be obtained also with a lowest RMS selection. However, such a method becomes sub-optimal once the observation cadence is not the same across all pulsars. For a more detailed investigation see Appendix~\ref{fig:appB_epta}}

%%%%%%%%%%%%%%%%%%%%%%%%%%%%%%%%%%%%%%%%%%%%%%%%%%%%%%%%%%%%%%%%%%%%%
%%%%%%%%%%%%%%%%%%%%%%%%%%%%%%%%%%%%%%%%%%%%%%%%%%%%%%%%%%%%%%%%%%%%%

%----------------------------------------------------------
\subsection{Optimizing IPTA and EPTA analysis of CGW signals}
%----------------------------------------------------------
We now test the performance of the CGW ranking method using noise-parameter values previously extracted from individual pulsar noise analyses of the latest IPTA data release \citep{2019MNRAS.490.4666P} and the {realistic} EPTA dataset created in the previous Sec.~\ref{subsec:optimize_epta}.
%%%%%%%%%%%%%%%%%%%%%%%%%%%%%%%%%%%%%%%%%%%%%%%%%%%%%%%%%%%%%%%%%%%%%
\begin{figure}
    \centering
    \includegraphics[width=0.5\textwidth]{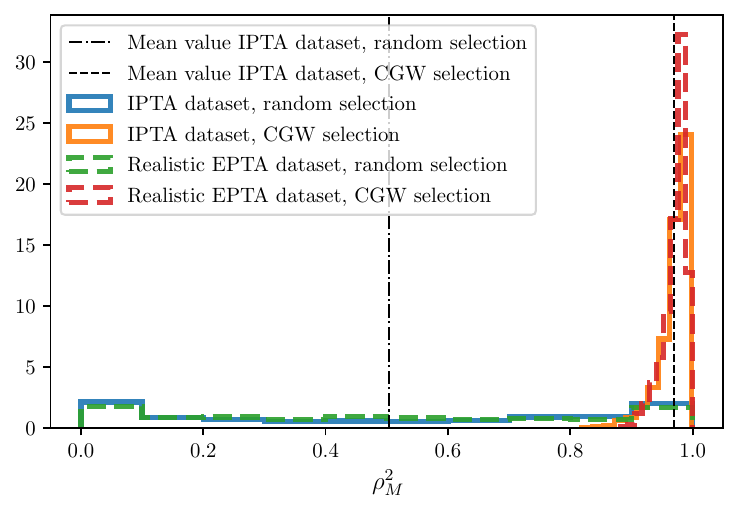}
    \caption{{Distribution of the normalized SNR$^2$} coverage for 1000 different sets of CGW parameters. The distributions are obtained with the list of pulsars {chosen according to the CGW selection method, in this case 22 for both the real IPTA dataset and the realistic EPTA dataset. For comparison, we also show the distribution of the normalized SNR$^2$ obtained with a random selection.}
    }
    \label{fig:snr_coverage}
\end{figure}
%%%%%%%%%%%%%%%%%%%%%%%%%%%%%%%%%%%%%%%

Because the ranking method is based on an exact noise-averaged formula, it is unnecessary to simulate noise realizations to test its performance.
However, we still want to prove that the selected pulsars recover most of the total SNR in the presence of a true (i.e. non-averaged) signal. We test this by comparing the fraction of total SNR$^2$ obtained using the CGW ranked pulsars to that obtained from a random pulsar selection. For an array of $N$ pulsars, the fraction of total SNR$^2$, given a list of $M < N$ pulsars, is defined as:
\begin{equation}
    \rho ^2 _M = \sum_{a=1}^M \bar{\rho} _a ^2 \, , \qquad \qquad \textrm{with } 0 < \rho_M ^2 < 1 \, ,
\end{equation}
where $ \bar{\rho} _a ^2$ is the normalized SNR$^2$ defined in Eq~(\ref{eq:norm_snr}).

After extracting the list of best pulsars, we test the selection procedure as follows:
\begin{itemize}

\item We draw the CGW signal parameters $\vec{\theta}$ from a uniform distribution with bounds defined as in the integral of Eq.~\eqref{eq:averaged_cw_snr_bounds}, and with frequency between 1 and 100 nHz. As pointed out in Sec.~(\ref{subsec:continuous_snr}) the strain amplitude has no influence on the ranking and therefore we fix it to $h=10^{-14}$.

\item We compute the non-averaged optimal SNR $\rho_\text{\tiny Opt} = \sqrt{(\vec{s}|\vec{s})} $ for each pulsar for a CGW signal $\vec{s}(t, \vec{\theta})$  and we use this quantity to calculate the normalized $ \bar{\rho} _a ^2$ defined in Eq.~(\ref{eq:norm_snr}). 

\item We compute $\rho ^2 _{M-\text{CGW}}$ for the list of best selected pulsars and $\rho ^2 _{M-\text{rand}}$ for a random subset of pulsars of random size $M$.

\item We repeat the previous steps one thousand times.

\end{itemize}

This gives us 1000 values of $\rho ^2 _{M-\text{CGW}}$ and $\rho ^2 _{M-\text{rand}}$ that we plot as histograms on Figure \ref{fig:snr_coverage}. For the IPTA dataset, the distribution of fractional $\rho ^2 _{M-\text{CGW}}$ for the selected pulsars is narrowly peaked around a mean value 0.97. The random selection $\rho ^2 _{M-\text{rand}}$ gives an almost uniform distribution with 0.50 mean value. The distribution is not uniform because $\rho _a ^2$ is not uniform and a few $\rho _a ^2$ values are much bigger while many others are very small. Similar results are obtained for the {realistic} EPTA dataset. We find that the number of pulsars which gives 95\% of the SNR$^2$ is 22 for both datasets{, and these pulsars represents respectively 61\% of the total number of TOAs ($=18584$) for the realistic EPTA dataset, and 76\%  of the total number of TOAs ($=210148$) for the IPTA dataset}.

Now we briefly discuss the comparison between the CGW and GWB selection methods.
Focusing on the realistic EPTA dataset, we find an overlap between the identified best pulsars with the CGW method and GWB method as shown in Table~\ref{tab:cgw_chim_snr}.
This time we run the Chimera and SNR$_\textrm{B}$-maximization ranking without fixing the six initial pulsars of the EPTA. We find that 17 pulsars are common to all three selection methods (highlighted in bold).

In summary, when true CGW signals are injected in the data, the CGW ranking method selects the pulsars which provides most of the SNR of the array, whereas a random selection is inefficient. This method extracts the few best pulsars to optimize the search for a CGW signal.
%%%%%%%%%%%%%%%%%%%%%%%%%%%%%%%%%%%%%%%%%%%%%%%%%%%%%%%%%%%%%%%%%%%%%%%%%%%%%%%%%%%%%%%%%%
\begin{table}
    \centering
    \begin{tabular}{|c|c|c|}
    CGW ranking & Chimera method & SNR$_\textrm{B}$ maximization\\ 
    \hline
\textbf{J0030+0451} & \textbf{J0030+0451} & \textbf{J0030+0451}\\ 
\textbf{J0613$-$0200} & J0034$-$0534 & \textbf{J0613$-$0200}\\ 
\textbf{J0751+1807} & \textbf{J0613$-$0200} & J0621+1002\\ 
J1012+5307 & J0621+1002 & \textbf{J0751+1807}\\ 
J1022+1001 & \textbf{J0751+1807} & J1022+1001\\ 
\textbf{J1024$-$0719} & J1012+5307 & \textbf{J1024$-$0719}\\ 
\textbf{J1600$-$3053} & \textbf{J1024$-$0719} & \textbf{J1600$-$3053}\\ 
\textbf{J1640+2224} & J1455$-$3330 & \textbf{J1640+2224}\\ 
\textbf{J1713+0747} & \textbf{J1600$-$3053} & \textbf{J1713+0747}\\ 
\textbf{J1730$-$2304} & \textbf{J1640+2224} & \textbf{J1730$-$2304}\\ 
\textbf{J1744$-$1134} & \textbf{J1713+0747} & \textbf{J1744$-$1134}\\ 
\textbf{J1751$-$2857} & \textbf{J1730$-$2304} & \textbf{J1751$-$2857}\\ 
\textbf{J1804$-$2717} & \textbf{J1744$-$1134} & J1801$-$1417\\ 
J1853+1303 & \textbf{J1751$-$2857} & \textbf{J1804$-$2717}\\ 
\textbf{J1857+0943} & J1801$-$1417 & J1843$-$1113\\ 
\textbf{J1909$-$3744} & \textbf{J1804$-$2717} & J1853+1303\\ 
\textbf{J1910+1256} & J1843$-$1113 & \textbf{J1857+0943}\\ 
J1911+1347 & \textbf{J1857+0943} & \textbf{J1909$-$3744}\\ 
\textbf{J1918$-$0642} & \textbf{J1909$-$3744} & \textbf{J1910+1256}\\ 
\textbf{J2010$-$1323} & \textbf{J1910+1256} & J1911+1347\\ 
\textbf{J2124$-$3358} & J1911$-$1114 & J1911$-$1114\\ 
J2145$-$0750 & \textbf{J1918$-$0642} & \textbf{J1918$-$0642}\\ 
  & \textbf{J2010$-$1323} & \textbf{J2010$-$1323}\\
  & \textbf{J2124$-$3358} & \textbf{J2124$-$3358}\\
  & J2322+2057 & J2322+2057\\
  \hline
  \end{tabular}
    \caption{List of the first 22 pulsars selected with the CGW ranking method and the 25 pulsars selected with the Chimera method and SNR$_\textrm{B}$-maximization in the realistic EPTA dataset. Bold font indicate the 17 pulsars that are selected by all three methods.}
    \label{tab:cgw_chim_snr}
\end{table}

%%%%%%%%%%%%%%%%%%%%%%%%%%%%%%%%%%%%%%%%%%%%%%%%%%%%%%%%%%%%%%%%%%%%%
%%%%%%%%%%%%%%%%%%%%%%%%%%%%%%%%%%%%%%%%%%%%%%%%%%%%%%

%----------------------------------------------------------
\section{Conclusions and future outlook}
\label{sec:conclusion}
%----------------------------------------------------------
PTA data analysis requires both significant human and computational resources. As the computational burden of such analyses grows with the number of pulsars, the problem will be further exacerbated by the discovery of many new pulsars by next-generation radio facilities. In this work, we introduced the concept of pulsar selection optimization for specific analyses. We emphasize that the ranking procedure is not straightforward and depends on the properties of the sought signal, and the optimization requirements. Therefore, we considered optimal selection criteria for deterministic CGW and stochastic GWB searches separately.

For the GWB, we presented three different ranking methods that target different aspects of a GWB search: SNR-maximization, Coupling Matrix, and Chimera method. The performance of our methods was assessed using frequentist and Bayesian hypothesis testing on simulated datasets.

The SNR-maximization method aims to increase the detection confidence in favor of the HD correlation with respect to a CURN process. Pulsars selected with this method provide an evidence for the HD vs CURN hypothesis larger than a random selection for all the considered datasets. For instance, using the EPTA-simplified dataset we obtained a log-Bayes factor which is double the one obtained with the random selection. Additionally, it was demonstrated that with this dataset we can reach 88\% of the total sensitivity after including $N=25$ pulsars out of $40$. The SNR-maximization method was further studied in Sec.~\ref{subsubsec:mock_epta} for the case of a {realistic} EPTA dataset with intrinsic red noise included.
We found that the first $\sim 20$ pulsars are included regardless of the particular noise realization and respective GWB parameter estimations. It was shown that the method selects pulsars which provide $1.8-1.9$ times larger log-likelihood ratio than a random selection. Furthermore, 25 pulsars out of the 40 selected by the SNR-maximization method accounted for 89\% of the log-likelihood ratio of the full dataset.

Inherently, the SNR-maximization method tends to pick pulsars that maximize the HD ORF, which results in clustering of the sources at angular separations of $0^\circ$ and $180^\circ$. This fact can be detrimental for disentangling the HD from other spatially correlated noise processes. The Coupling Matrix selection is aimed at resolving this issue by maximizing the decoupling between different correlations, so that the HD spatial mode disentangles from the monopolar and dipolar correlations. This method has been shown to be efficient at increasing the evidence in the hypothesis test HD vs CURN+MN+DN in two out of the three datasets. The main pitfall of this method is that it weakly depends on the relative sensitivity of selected sources. As a consequence, some of the high-SNR sources are left behind, which is the main reason for the loss of sensitivity to GWB.

The Chimera method combines the two approaches to optimize both the sky coverage and the gain in total SNR. Even though its formulation is heuristic, this selection method has been a good proxy for selecting the pulsars that increase confidence in a GWB detection comparable to Coupling Matrix and SNR maximization. Specifically, for the simplified-EPTA dataset the method is able to recover 90\% of the sensitivity of the whole array with $N=25$ pulsars. In future work this formalism is going to be further examined. In particular, it would be interesting to explore if the Information matrix formalism introduced recently in \cite{Ali-Haimoud:2020iyz,Ali-Haimoud:2020ozu} could be used to develop a more rigorous Chimera method, or a selection method targeting anisotropic searches.

The CGW SNR maximization is constructed to find the best pulsars to detect a CGW from a SMBHB. In contrast to the GWB case, CGW ranking deals with purely deterministic signals and this allows us to treat every pulsar independently, within our formalism. The method is based on an averaged SNR formula, and was applied to {continuous wave} signal searches in the IPTA and {realistic} EPTA mock datasets. Because of the strong dependence of an individual pulsar's SNR response $\bar{\rho} _a (f)$ on the CGW frequency $f$, ranking was performed separately for different frequency bins. In order to find the best pulsars on some frequency range, we had to take the union of the best pulsars that were identified for several frequency bins. Using the 22 best-ranked pulsars we recovered more then 95\% of the total SNR$^2$ for both the IPTA and {realistic} EPTA datasets.
Furthermore, we found that 17 of these pulsars are also selected by the SNR-maximization and Chimera methods.

The main takeaway points of our study can be summarized as follows:
\begin{itemize}
\item Although the addition of new pulsars inevitably increases the sensitivity of a PTA towards CGW and GWB detection \citep[see][]{Siemens_2013}, there exists an optimal subset of pulsars which is responsible for a larger portion of the sensitivity of a PTA, especially if the pulsar have different noise properties. This behaviour is confirmed in Figure \ref{fig:cumul_plot} for CGWs, and Figures \ref{fig:simpli_epta_dataset} and \ref{fig:lnL_comparison} for a GWB. If pulsars have all equal noise properties, it is possible to include pulsars such that the increase in the evidence is steeper than a random selection. This can be seen in Figure~\ref{fig:galaxy_dataset}.

\item In contrast to intuitive expectations, covering the sky uniformly with pulsars is not the most optimal strategy of pulsar selection for the purpose of disentangling different spatial modes, even in the case that all pulsars are equally sensitive. Instead, as can be seen from Figure~\ref{fig:ang_distr}, the ultimate distribution of pulsars in $\cos \theta_{a b}$ has three distinctive peaks at angular separations of $0^\circ, 90^\circ$ and $180^\circ$. We expect that this distribution will converge to a uniform distribution, if we aim to resolve all multipoles.

\item We stress that although a high SNR provides a steeper increase in the log-Bayes factor when HD is compared to all other considered types of common processes, it does not guarantee an optimal decoupling of spatial modes. This is clearly illustrated with the Galaxy-distributed and mock MeerTime datasets.

\item {Good sky coverage alone does not guarantee the effective decoupling of spatial modes. The optimal pulsar selection criterion should balance between proper sky localization and high sensitivity. The Chimera method is an attempt to create such a criterion which accounts for both properties. However, as demonstrated in Appendix~\ref{subsec:rms_selection}, simpler selection methods might perform better than the Chimera method for some datasets. The optimal weighting between the position and the sensitivity of a pulsar will be the subject of future investigations.}

\end{itemize}

The purpose of these ranking methods is not to discard the analysis of some pulsars but only to evaluate their contribution to the full PTA analysis. Even though these results depend on the noise properties of the PTA dataset considered, the selection of a subset of pulsars has been shown to be a good proxy for having an informative dataset and at the same time reducing the computational burden of the analysis.
Therefore, if a collaboration decides to limit pulsar sources due to resource restrictions, these tools will be essential for understanding how to make such a selection. 
These methods will be crucial to extend the array of existing experiments and target specific analyses when the next generation of radio facilities discover a large number of new pulsars.

%----------------------------------------------------------
\section*{Acknowledgements}
%----------------------------------------------------------

We thank Stanislav Babak, Golam Shaifullah, Anuradha Samajdar, David Champion, Aditya Parthasarathy for useful discussions. {We are very thankful to the anonymous referee for improving this manuscript.}
SRT acknowledges support from NSF AST-2007993, the NANOGrav NSF Physics Frontier Center PHY-2020265, and an NSF CAREER Award PHY-2146016.
AS acknowledges financial support provided under the European Union’s H2020 ERC Consolidator Grant ``Binary Massive Black Hole Astrophysics'' (B Massive, Grant Agreement: 818691).
We made use of numpy and scipy \citep{2020SciPy-NMeth,harris2020array}.

%%%%%%%%%%%%%%%%%%%%%%%%%%%%%%%%%%%%%%%%%%%%%%%%%%
\section*{Data Availability}
The timing data and codes used in this article shall be shared on reasonable request to the corresponding authors.



\bibliographystyle{mnras}
\bibliography{ranking_pls.bib} % if your bibtex file is called example.bib

%%%%%%%%%%%%%%%%%%%%%%%%%%%%%%%%%%%%%%%%%%%%%%%%%%

%%%%%%%%%%%%%%%%% APPENDICES %%%%%%%%%%%%%%%%%%%%%

\appendix

\section{Implementation of different weights for Coupling Matrix formalism optimization}
\label{subsec:cm_wgt}
In this paragraph we provide further clarifications on the choice of the weighting function $w_{\alpha}$ from Eq.(\ref{eq:cp_wght}). As mentioned in the main text, the weights for the construction of the coupling matrix should have a direct correspondence to the relative sensitivity of a source in an array. Here, we tested the performance of the Coupling Matrix formalism using as the weighting function $\textrm{SNR}_\text{A}$ raised to the power of 2, 4 and 6. The optimal performance is obtained using $\textrm{SNR}^4_\text{A}$-weights. Coupling matrix selection with weights of lower power of $\textrm{SNR}_\text{A}$ tends to pick pulsars with a triple-peak distribution on the sky (see Figure \ref{fig:ang_distr}), while the individual sensitivity of a source is relegated to the background. The degradation of the efficiency of $\textrm{SNR}^6_\text{A}$-weighting for the mock MeerTime dataset is due to a saturation of the coupling matrix by the high SNR pulsars, so that it becomes essentially insensitive to adding further sources of lower sensitivity, or in some cases even ill-defined. In order to evade the problem of saturation, we have proposed to use the eigenvalue-ratio $\delta_\lambda$ ($w^a=1$) and the individual SNRs of the pulsars combined in a Chimera-like manner: $\delta_\lambda \prod_{a=1}^{N_{\textrm{psr}}}\textrm{SNR}^a_\textrm{A}$. The performance of the latter method is comparable to the one of the Coupling Matrix formalism with $\textrm{SNR}^4_\text{A}$-weights. The efficacy of the Coupling Matrix selection and its modifications is going to be investigated more thoroughly in future work on a broader range of datasets.
\begin{figure}
    \centering
    \includegraphics[width=0.5\textwidth]{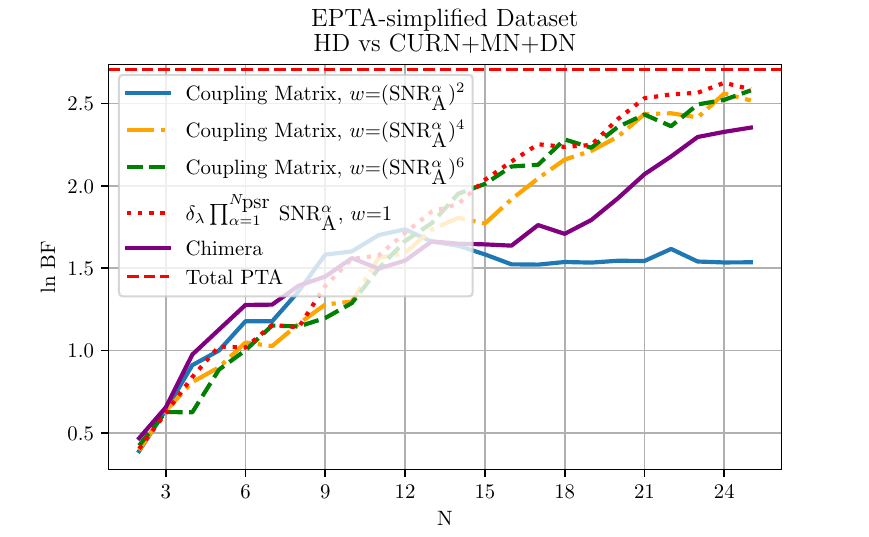}
    \includegraphics[width=0.5\textwidth]{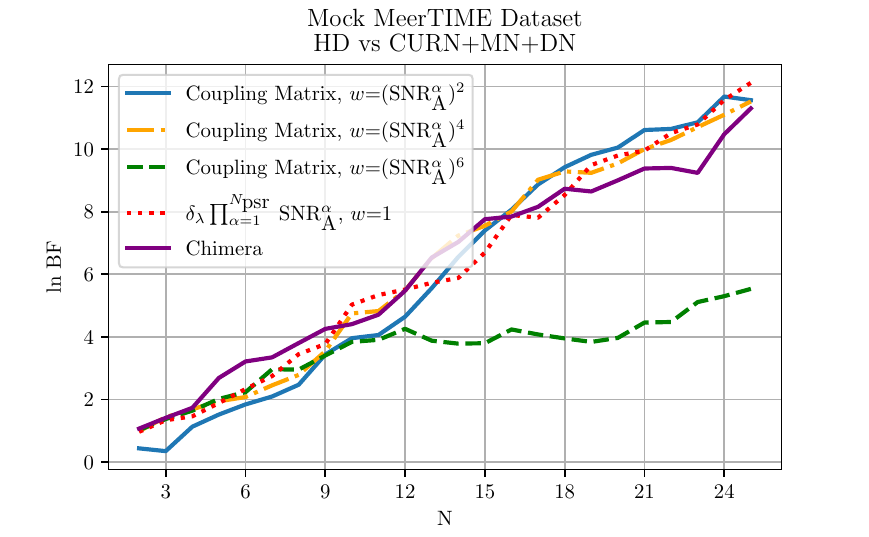}
    \caption{
    Log-Bayes factor of the hypothesis test HD vs CURN+MN+DN as a function of the number of pulsars selected by various modifications of the Coupling Matrix formalism (shown in different colors). The corresponding result for the Chimera method (purple color) are also shown for comparison. The upper panel shows the result for the simplified EPTA dataset {averaged over 45 noise realizations}, and the log-Bayes factor of the full array is indicated with a horizontal red dashed line. The bottom panel demonstrates the results for the mock MeerTime dataset.
    }
    \label{fig:cm_weight}
\end{figure}

\section{{Simple alternative selection methods}}
\label{subsec:rms_selection}
Throughout the paper we compared our selection methods to a random pulsar selection, because only a random selection can be considered independent of the specifics of the datasets. However, such a selection method would not be adopted in a realistic setting. Therefore, we explore how the selection methods compare to more realistic, still simple, ranking criteria: selecting pulsars based on their lowest RMS noise and longest timespan.

For the case of the Galaxy-distributed dataset (Sec.~\ref{subsubsec:galaxy}) where all the pulsars have the same RMS and timespan, it is already clear that our ranking methods outperform a lowest-RMS selection or a longest-timespan selection, which are equivalent to the random selection. For the EPTA-simplified dataset (Sec.~\ref{subsubsec:mock_epta}) and the Mock MeerTime dataset (Sec.~\ref{subsubsec:meertime}) we perform only the RMS selection because all the pulsars' timespans are equal.

\begin{figure}
    \centering
    \includegraphics[width=0.5\textwidth]{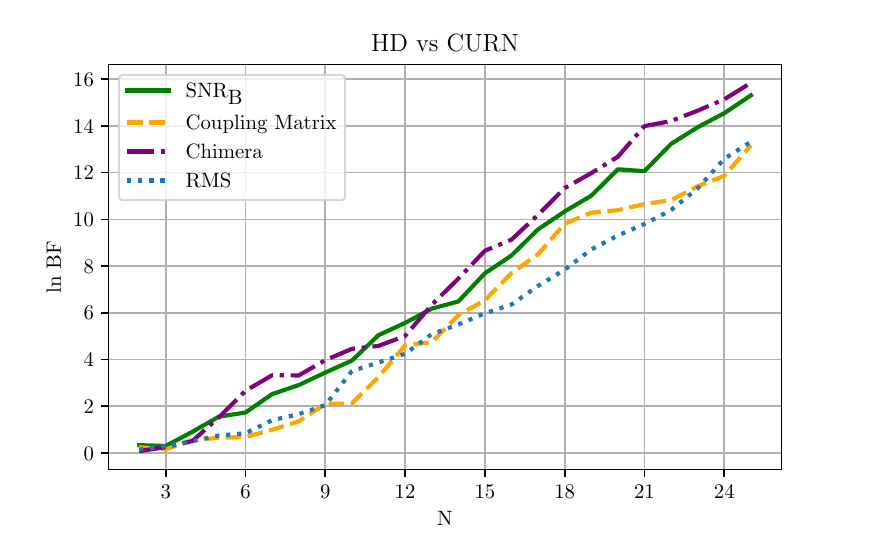}
    \includegraphics[width=0.5\textwidth]{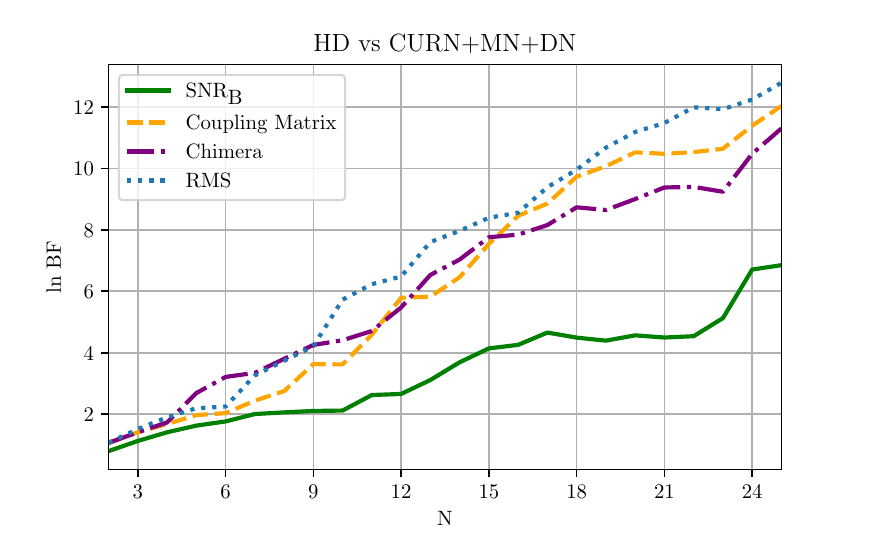}
    \caption{Log-Bayes factor as a function of the number of chosen pulsars for each of the selection methods (shown in different colors) for the Mock MeerTime dataset and for different hypothesis tests: HD vs CURN (top), and HD vs CURN+MN+DN (Bottom). The shown log-Bayes factors represent the average over 20 different noise realizations.
    }
    \label{fig:appB_meertime}
\end{figure}

For the Mock MeerTime dataset (Fig.~\ref{fig:appB_meertime}), the RMS selection method provides Bayes factors comparable to those of the Coupling Matrix and worse than the $\textrm{SNR}_\text{B}$ and Chimera method, for the hypotheis test HD vs CURN. However, for the hypothesis test HD vs CURN+MN+DN, the RMS selection method performs better than all the others.

\begin{figure}
    \centering
    \includegraphics[width=0.5\textwidth]{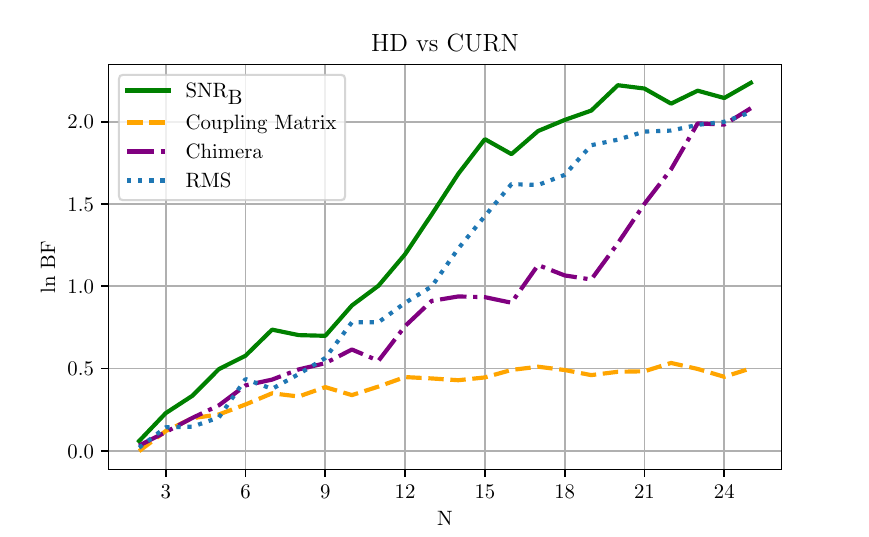}
    \includegraphics[width=0.5\textwidth]{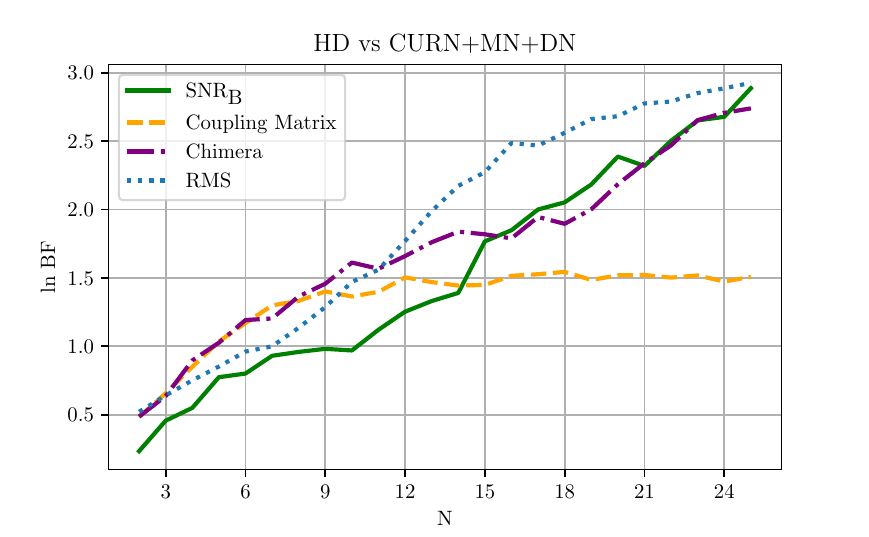}
    \caption{Log-Bayes factor as a function of the number of chosen pulsars by each of the selection methods (shown in different colors) for the EPTA-simplified dataset and for different hypothesis tests: HD vs CURN (top), and HD vs CURN+MN+DN (Bottom). The shown log-Bayes factors represent the average over 20 different noise realizations.
    }
    \label{fig:appB_epta}
\end{figure}

For the EPTA-simplified dataset (Sec.~\ref{subsubsec:mock_epta}) the results are shown in Fig.~\ref{fig:appB_epta}. The RMS selection method provides Bayes factors comparable to the ones of the Chimera method for 25 pulsars and slightly smaller than the $\textrm{SNR}_\text{B}$ method, for the hypothesis test HD vs CURN. For the hypothesis test HD vs CURN+MN+DN, the RMS selection method yields a Bayes factor comparable to the one of the $\textrm{SNR}_\text{B}$ selection. 
The reason why for the hypothesis test HD vs CURN+MN+DN in the EPTA-simplified and mock MeerTime datasets the RMS selection performs better than other selection methods is that the lowest-RMS pulsars are almost uniformly distributed on the sky, so that the most sensitive pulsars of the array are picked in sufficiently optimal parts of the sky. For the arrays in which low-RMS pulsars are clustered in a specific region of the sky, this will not be the case. For the hypothesis test HD vs CURN, the RMS method does not differ significantly from the SNR-maximization, because the SNR formula already takes into account the RMS values and the aforementioned datasets are affected only by white noise.

\begin{figure} 
\includegraphics[width=1.0\columnwidth]{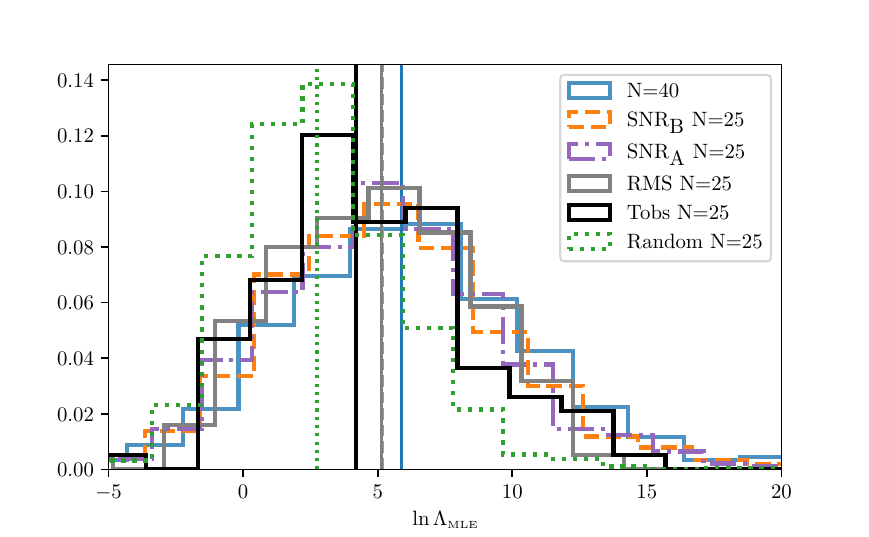} 
\includegraphics[width=1.0\columnwidth]{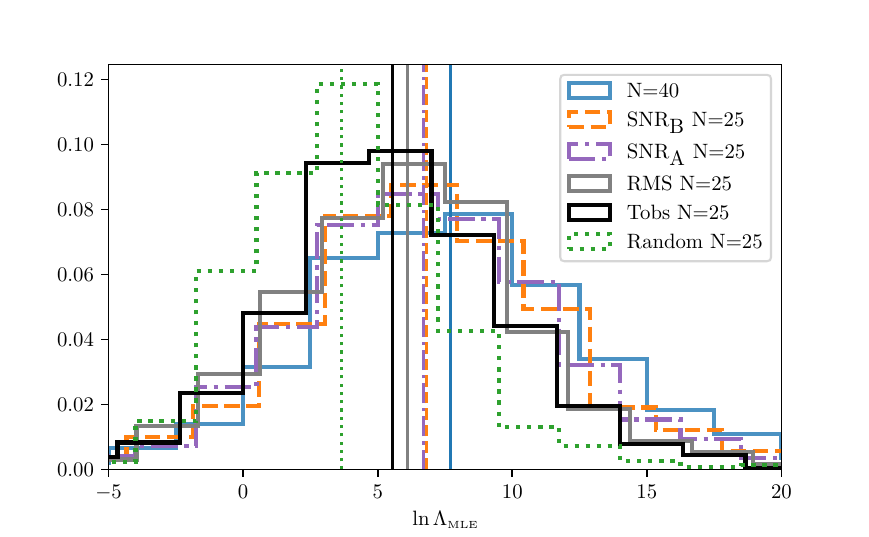} 
\caption{(\emph{Top}): Distribution of log-likelihood ratios obtained as in Fig.~\ref{fig:lnL_comparison} but with the addition of the distributions of log-likelihood ratios obtained with the lowest-RMS (RMS) and the longest-timespan (Tobs) selections. The median values for the shown distributions are: 5.88 (N=40), 5.17 ($\text{SNR}_{\rm B}$), 5.14 ($\text{SNR}_{\rm A}$), 5.13 (RMS), 4.19 (Tobs), 2.73 (Random). (\emph{Bottom}): Same analysis as above but for the simulated realistic EPTA dataset with a number of TOAs as in the real EPTA dataset and not every 14 days as in the (simulated) realistic EPTA dataset.
The median values for the shown distributions are: 7.71 (N=40), 6.80 ($\text{SNR}_{\rm B}$), 6.69 ($\text{SNR}_{\rm A}$), 6.11 (RMS), 5.53 (Tobs), 3.66 (Random).
}
\label{fig:appB_loglike}
\end{figure}

For the realistic EPTA datasets (Sec.~\ref{subsec:optimize_epta}), we performed the lowest-RMS and longest-timespan selections, and we show the results in the top panel of Fig.~\ref{fig:appB_loglike}. The lowest-RMS selection does not seem to differ from the SNR-maximization selection and it yields in median approximately the same log-likelihood ratio, which is $\sim 0.87$ times the total one. The longest-timespan selection performs slightly worse than the SNR-maximization and lowest-RMS selections, and it provides a log-likelihood ratio $0.71$ times the one from the full dataset.

To highlight the difference between the lowest-RMS selection and the SNR-maximization selection we created a new dataset which is identical to the realistic EPTA dataset of Sec.~\ref{subsec:optimize_epta}, apart from the number of TOAs of each pulsar. The pulsars simulated for the realistic EPTA dataset have the same timespan as the real EPTA dataset, but with TOAs observed every 14 days. Now, the new dataset has the same number of TOAs as the real EPTA dataset and their TOA cadence range between one per day up to one every 18 days. The results of the same analysis of Sec.~\ref{subsec:optimize_epta} are shown in the bottom panel of Fig.~\ref{fig:appB_loglike}. Contrary to the previous results, the lowest-RMS selection method is now sub-optimal compared to the SNR-maximization method. The contribution to the total noise power due to white and red noise has changed as the TOA cadence is different. This has an impact on the selection methods. In fact, the SNR ranking recovers 88\% of the total log-likelihood, whereas the lowest-RMS selection reaches only 79\%.

Even if the SNR-ranking method does not perform as well as the RMS selection in some scenarios, it is more flexible and its relatively cheap computational cost makes it worth using it instead of RMS or longest timespan selection, when testing the HD vs CURN hypothesis.

% Don't change these lines
\bsp	% typesetting comment
\label{lastpage}
\end{document}